\documentclass{aa}  
\usepackage{graphicx}
\usepackage{lscape}
\usepackage[varg]{txfonts}
\usepackage{xspace}
\usepackage{amssymb}
\usepackage{stmaryrd}
\usepackage{color}
\usepackage[usenames,dvipsnames]{xcolor}
\usepackage{natbib}
\usepackage{url}
\bibpunct{(}{)}{;}{a}{}{,}

\newcommand{\hsa}{\object{HS\,1522+6615}\xspace}
\newcommand{\hsb}{\object{HS\,2209+8229}\xspace}
\newcommand{\hsc}{\object{HS\,0111+0012}\xspace}
\newcommand{\ic}{\object{IC\,4663}\xspace}
\newcommand{\ks}{\object{KS\,292}\xspace}
\newcommand{\gjjc}{\object{GJJC\,1}\xspace}
\newcommand{\lsea}{\object{LSE\,153}\xspace}
\newcommand{\lseb}{\object{LSE\,263}\xspace}
\newcommand{\lsec}{\object{LSE\,259}\xspace}
\newcommand{\pnk}{\object{K\,1$-$27}\xspace} 
\newcommand{\kpd}{\object{KPD\,0005+5106}\xspace}
\newcommand{\pnl}{\object{LoTr\,4}\xspace}
\newcommand{\wda}{\object{PG\,0038+199}\xspace}
\newcommand{\wdb}{\object{PG\,1034+001}\xspace}
\newcommand{\wdc}{\object{PG\,0108+101}\xspace}

\newcommand{\abe}{\object{Abell\,48}\xspace}

\newcommand{\cci}{\object{PG\,0046+078}\xspace} 
\newcommand{\ccj}{\object{PG\,0237+116}\xspace}

\newcommand{\kwa}{\object{SDSS\,J\,171916.97+365326.70}\xspace} 
\newcommand{\kwb}{\object{SDSS\,J\,141812.50+024426.92}\xspace} 
\newcommand{\kwc}{\object{SDSS\,J\,075732.18+184329.28}\xspace}

\newcommand{\kwn}{\object{SDSS\,J\,172854.34+361958.62}\xspace}

\newcounter{Rco}

\newcommand{\Ion}[2]{\ion{#1}{#2}}
\newcommand{\Ionw}[3]{\mbox{\ion{#1}{#2}~$\lambda\,#3$\,\AA}}

\newcommand{\Ionww}[3]{\mbox{\ion{#1}{#2}~$\lambda\lambda\,#3$\,\AA}}

\newcommand{\logg}{\mbox{$\log g$}}
\newcommand{\loggw}[1]{\mbox{$\log g\hspace{-0.5mm} =\hspace{-0.5mm}  #1$}}

\newcommand{\se}[1]{\mbox{Sect.\,\ref{#1}}}

\newcommand{\sK}[1]{\mbox{(Sect.\,\ref{#1})}}

\newcommand{\Teff}{\mbox{$T_\mathrm{eff}$}}
\newcommand{\Teffw}[1]{\mbox{$\Teff\hspace{-0.5mm} =\hspace{-0.5mm} #1 \,\mathrm{kK}$}}
\newcommand{\ebv}{\mbox{$E_{B-V}$}\xspace}

\newcommand{\Msol}{$M_\odot$}

\begin{document}

\title{On helium-dominated stellar evolution: \\
       the mysterious role of the O(He)-type stars
        \thanks
        {Based on observations with the NASA/ESA Hubble Space Telescope, obtained at the Space Telescope Science 
         Institute, which is operated by the Association of Universities for Research in Astronomy, Inc., under 
         NASA contract NAS5-26666.
        }$^{,}$
        \thanks
        {Based on observations made with the NASA-CNES-CSA Far Ultraviolet Spectroscopic Explorer.
        }$^{,}$
        \thanks
        {Based on observations made with ESO Telescopes at the La Silla Paranal Observatory under programme IDs 091.D-0663, 090.D-0626}
        }
\author{N\@. Reindl\inst{1}
        \and
        T\@. Rauch\inst{1}
        \and
        K\@. Werner\inst{1}
        \and
        J\@. W\@. Kruk\inst{2}
        \and 
        H\@. Todt\inst{3}}

\institute{Institute for Astronomy and Astrophysics,
           Kepler Center for Astro and Particle Physics,
           Eberhard Karls University, 
           Sand 1,
           72076 T\"ubingen, 
           Germany\\
           \email{reindl@astro.uni-tuebingen.de}
           \and       
           NASA Goddard Space Flight Center, Greenbelt, MD\,20771, USA
           \and       
           Institute for Physics and Astronomy, 
           University of Potsdam, 
           Karl-Liebknecht-Str\@. 24/25, 
           14476 Potsdam, 
           Germany}

\date{Received January 2014; accepted 2014}

\abstract{About a quarter of all post-asymptotic giant branch (AGB) stars are hydrogen-deficient.
          Stellar evolutionary models explain the carbon-dominated H-deficient stars 
          by a (very) late thermal pulse scenario 
          where the hydrogen-rich envelope is mixed with the helium-rich
          intershell layer. Depending on the particular time at which the
          final flash occurs, the entire hydrogen envelope may be burned. 
          In contrast, helium-dominated post-AGB stars 
          and their evolution are yet not understood.}
          {A small group of very hot, helium-dominated stars is formed by O(He)-type stars. A precise analysis of their
           photospheric abundances will establish constraints to their evolution.}
          {We performed a detailed spectral analysis of ultraviolet and optical spectra 
           of four O(He) stars by means of state-of-the-art non-LTE model-atmosphere techniques.}
          {We determined
           effective temperatures, 
           surface gravities, 
           and the abundances of H, He, C, N, O, F, Ne, Si, P, S, Ar, and Fe. 
           By deriving upper limits for the mass-loss rates of the O(He) stars, we found that they 
           do not exhibit enhanced mass-loss. 
           The comparison with evolutionary models shows that the
           status of the O(He) stars remains uncertain.
           Their abundances match predictions of a double helium white dwarf merger
           scenario, suggesting that they might be the progeny of the compact and of the luminous 
           helium-rich sdO-type stars. The existence of planetary nebulae that do not show helium enrichment 
           around every other O(He) star, precludes a merger origin for these stars. 
           These stars must have formed in a different way, for instance via enhanced mass-loss during their post-AGB evolution 
           or a merger within a common-envelope (CE) of a CO-WD and a red giant or AGB star.
           }
          {A helium-dominated stellar evolutionary sequence exists, that may be
           fed by different types of mergers or CE scenarios. It appears likely, that all these
           pass through the O(He) phase just before they become white dwarfs.
           }


\maketitle

\section{Introduction}
\label{sect:introduction} 

Quantitative spectral analyses of hot, post-asymptotic giant branch (AGB) stars revealed two distinct evolutionary sequences. 
Besides the well understood H-rich sequence, a H-deficient sequence was discovered. 
It is composed of Wolf-Rayet-type stars that evolve into PG\,1159-type stars and finally might evolve into 
non-DA white dwarfs (WDs). While (very) late thermal pulse ((V)LTP) evolutionary models can explain the 
observed He, C, and O abundances in these stars \citep[the typical abundance pattern for PG\,1159 stars is 
He\,:\,C\,:\,O = 0.30\,$-$\,0.85\,:\,0.15\,$-$\,0.60\,:\,0.02\,$-$\,0.20 by mass,][]{wernerherwig2006}
they do not reproduce the abundances in He-dominated stars, such as subdwarf O (sdO) stars, R Coronae Borealis (RCB) stars, 
extreme helium (EHe) stars, and the O(He) stars.

Two evolutionary scenarios for the origin of RCB and EHe stars were suggested. They might either be formed 
by a final He-shell flash or be the merger product of a CO WD with a He WD \citep{jefferyetal2011}. The significant Li
content in the atmospheres of RCB stars supports the idea of an LTP to explain their origin.
However, the relatively high inferred masses of RCB stars and their high F abundance
supports a WD merger \citep{claytonetal2011}. \cite{zhangetal2012b} suggested that a double He-WD 
merger might also explain RCB and EHe stars.
Recently, \citet{rao2013} found the RCB star \object{DY\,Cen} to be the first and only binary system among the RCB stars 
and their probable relatives. \object{DY\,Cen} is one of the hottest and most H-rich members of the RCB stars. \citet{rao2013} 
suggested that this system might have evolved from a common-envelope system to its current form. Therefore it may be
possible that RCB stars form in various ways.

RCB and EHe stars are not the only He-rich stars for which a merger origin was suggested. \cite{zhangetal2012a} 
presented the results of a double He-WD merger to explain the formation of He-rich, 
hot sdOs. These can be divided into three subgroups, one C-rich and N-poor, the other N-rich and C-poor, and the 
third one enriched in C and N \citep{hirsch2009}. In their numerical experiments, \citet{zhangetal2012a} 
showed that in terms of \Teff, \logg, C, and N abundances, the origin of the two sdO groups can be explained by 
different double He-WD merger models. \citet{zhangetal2012a} distinguished between three types of mergers. In a 
slow-merger process the less massive He WD transfers its entire mass within a few minutes to form a disk 
around the primary He WD, which then accretes from it at 
a rate similar to the Eddington-accretion rate. The surface composition of the resulting star 
retains the N-rich composition of the accreted WD. In the fast-merger model, the secondary 
directly transfers its entire mass quickly to the primary surface, where heating up to $10^8$\,K 
causes the material to expand and form a hot corona within a few minutes. The fast-merger model produces 
C-rich stars, in which N is hardly observable. The composite merger includes both processes. 
In general, models with a merged mass below $0.7\,M_\odot$ produce an 
N-rich star, while more massive ones produce a C-rich star with a significant N content.\\
Other origins of He-rich sdO stars are discussed as well. 
Because the merger of a He WD with a post-sdB star is predicted in one of the important binary channels for the formation 
of sdB stars \citep{han2002, han2003}, \cite{justham2011} proposed this formation channel for the previously 
unexplained He-rich sdO stars. Their models also reproduced the properties of the He-rich sdO stars in terms of 
\Teff\ and \logg. However, in contrast to \citet{zhangetal2012a}, they did not include nuclear evolution 
in their calculations. 

A more exotic group of He-dominated stars is formed by the O(He) stars. This spectroscopic 
class is defined by an almost pure \Ion{He}{II} absorption-line spectrum in the optical
wavelength range \citep{mendezetal1986,rauchetal1998}. For more than 15 years, only the two central stars of
planetary nebulae (CSPNe) \pnk\footnote{In the following, we use the PNe names to identify their central stars.} (\object{WD\,0558$-$756}) 
and \pnl, as well as \hsa (\object{WD\,1522+662}), and \hsb (\object{WD\,2209$-$8229}) were known 
\citep{rauchetal1994, rauchetal1996, rauchetal1998}. Recently, \citet{werneretal2014} found four more of these objects, 
namely \kwn, \kwa, \kwb, and \kwc. They also pointed out that \kpd is a pre-WD. Because of the He-rich surface composition of 
\kpd (98\% by mass, \citealt{wassermannetal2010}), we also consider it as an O(He) star.
The evolutionary status of these objects has been studied only rarely. \cite{rauchetal1998} proposed that O(He) stars 
might be succesors of the luminous He-rich sdO stars and that because of due ongoing mass loss, the low-gravity O(He) stars might envolve 
into PG\,1159 stars. This possibility was studied by \cite{millerbertolamialthaus2006}. They had to assume mass-loss rates, that were higher 
than predicted by radiation-driven wind theory, to turn O(He) stars into helium-enriched PG\,1159 stars. An alternative scenario 
was suggested by \cite{rauchetal2006}. They invoked the idea that the O(He) stars could also be the offspring of a merging event 
of two white dwarfs and thus, the direct descendants of RCB stars.\\
In this paper, we present the analysis of \pnk, \pnl, \hsa, and \hsb. In \se{sect:observation}, we briefly describe 
the available observations. We then present a detailed spectral analysis of the O(He) stars based on optical and ultraviolet
spectra \sK{sect:analysis} and derive stellar parameters and distances \sK{sect:masses}. In \se{sect:discussion} 
we discuss possible evolutionary channels for the O(He) stars and those of other helium-dominated objects. We conclude in \se{sect:conclusions}.

\section{Observations}
\label{sect:observation} 

In January, June and July 2013, we performed observations of \pnk and \pnl with the ESO/NTT 
(ProgIDs 091.D-0663, 090.D-0626) using the EFOSC2 spectrograph (resolving power 
$R=\lambda / \Delta \lambda \approx 3000$) with the grisms GR$\#$19 (4400 to 5100\,\AA) and 
GR$\#$20 (6000 to 7100\,\AA). The data reduction was done by using 
IRAF\footnote{IRAF is distributed by the National Optical Astronomy Observatory, which is 
operated by the Associated Universities for Research in Astronomy, Inc., under cooperative 
agreement with the National Science Foundation.}.
The previous optical spectra of the O(He) stars were already described and analyzed by 
\citet{rauchetal1994,rauchetal1996,rauchetal1998} 
to determine effective temperature (\Teff), surface gravity (\logg), and the H/He ratio 
for all four stars as well as the C abundance for \pnk, \pnl (upper limits), \hsa, and 
the N abundance for \pnk and \pnl. 

We have taken FUV spectra with FUSE\footnote{Far Ultraviolet Spectroscopic Explorer} 
($R \approx 20000$) in 2002 (all O(He) stars, 49\,ksec exposure time in total, ProgID: C178) and 
2005 (only \hsa, 4\,ksec, ProgID: U103)
using the LWRS aperture (Figs.~\ref{fig:fuse1a} $-$ \ref{fig:fuse4b}). Additional 
204\,ksec of FUSE observations (\pnk, \hsa, \hsb) were scheduled for summer 2007,
but were not be performed because of the FUSE failure on July 12.
The FUSE spectra show a strong contamination by interstellar (IS) line absorption 
and the S/N ratio is very poor \citep{rauchetal2006,rauchetal2009}. To reduce the pixel-to-pixel 
variation, they were co-added and then slightly smoothed with a low-pass filter \citep{savitzkygolay1964}. 

\onlfig{
\begin{landscape}
\addtolength{\textwidth}{6.3cm} 
\addtolength{\evensidemargin}{-3cm}
\addtolength{\oddsidemargin}{-3cm}
\begin{figure*}
\includegraphics[trim=0 0 0 -120,height=20.0cm,angle=0]{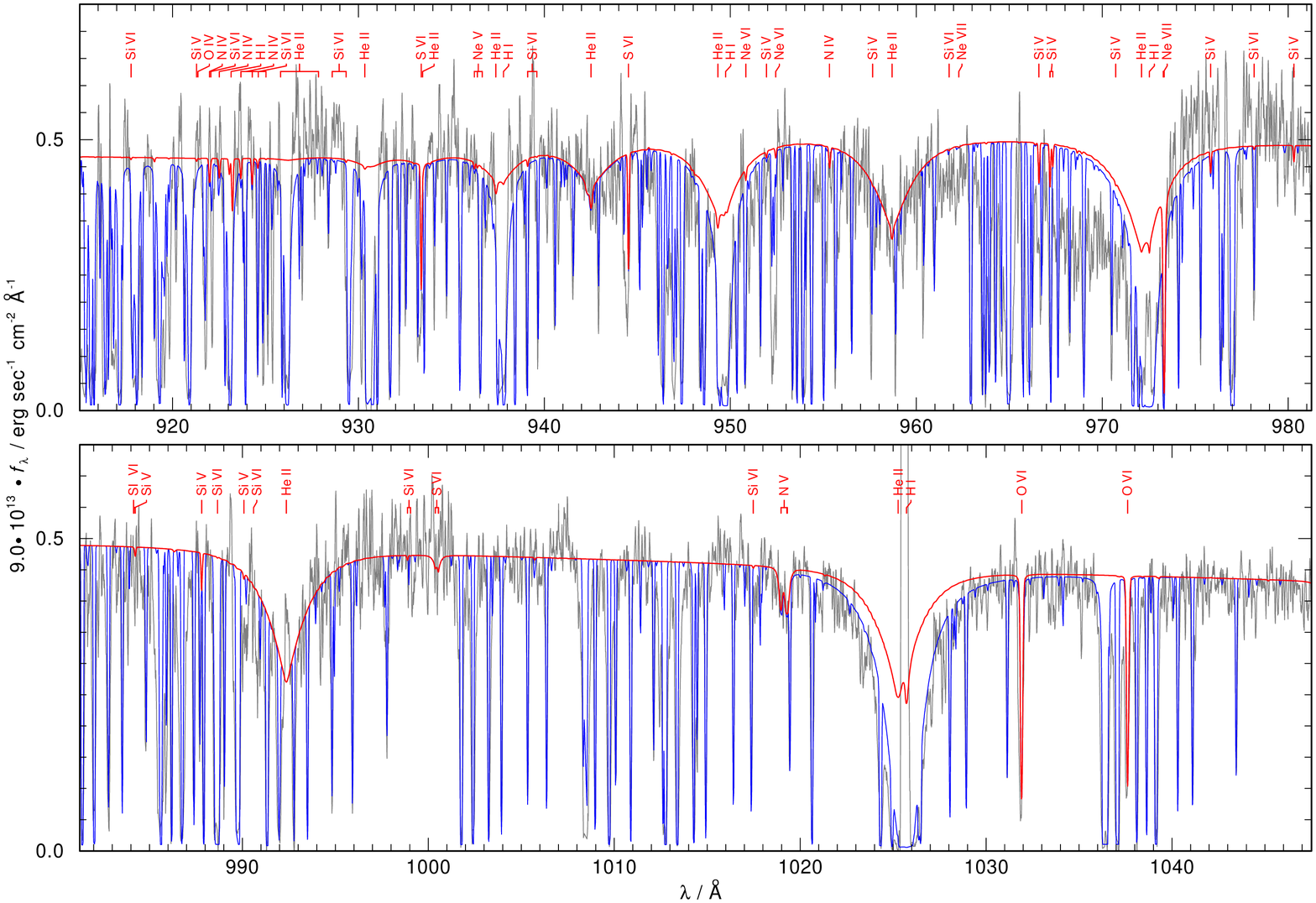}
  \caption{FUSE spectrum of \pnk (gray) compared with final synthetic spectra (red: pure stellar, 
           blue: combined stellar and interstellar). The locations of photospheric lines appearing in the synthetic spectrum are marked.}
  \label{fig:fuse1a}
\end{figure*}
\end{landscape}
}
  
  \addtocounter{figure}{-1}
  
  \onlfig{
  \begin{landscape}
  \addtolength{\textwidth}{6.3cm} 
  \addtolength{\evensidemargin}{-3cm}
  \addtolength{\oddsidemargin}{-3cm}
  \begin{figure*}
    \includegraphics[trim=0 0 0 -120,height=20.0cm,angle=0]{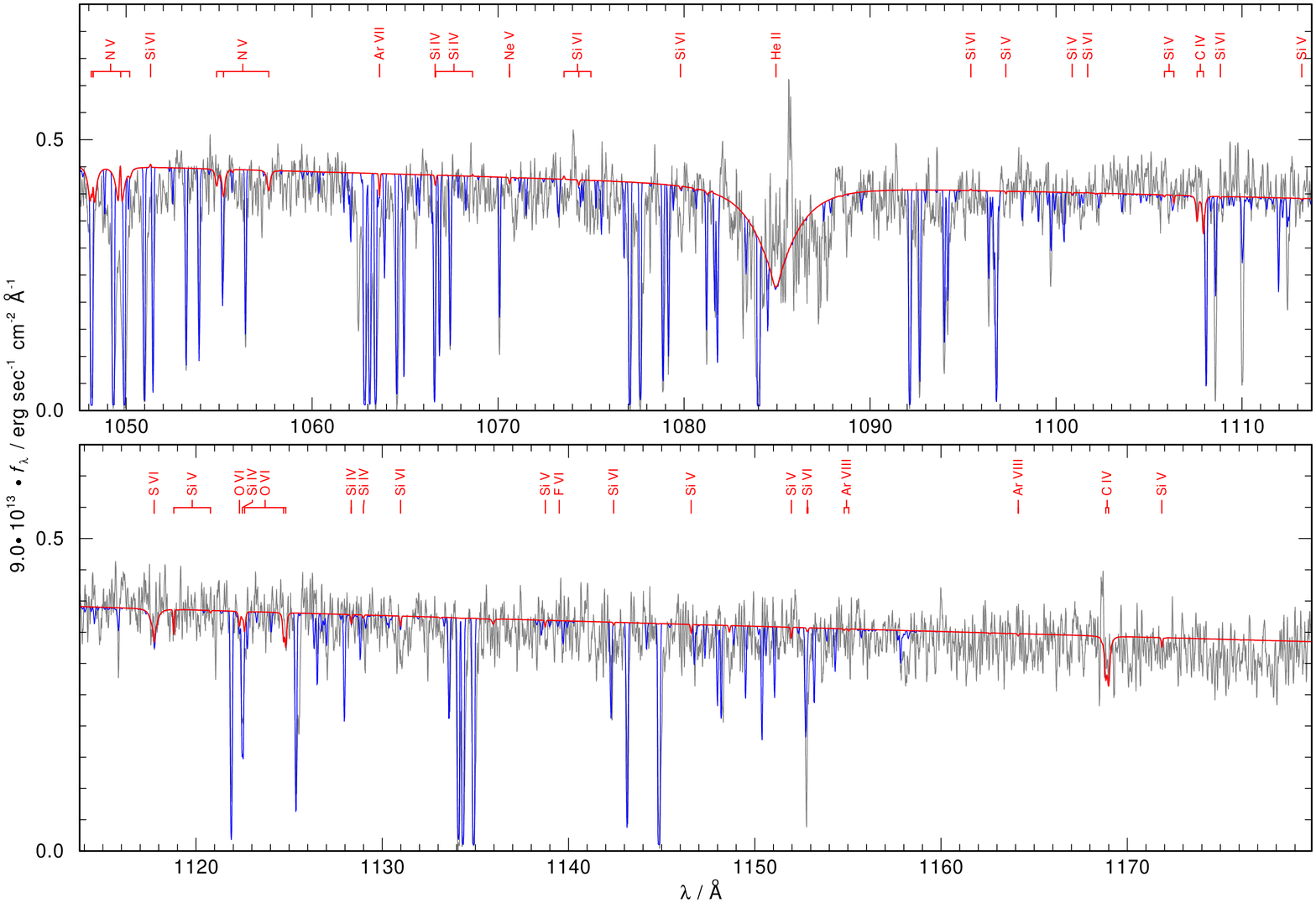}
    \caption{Continued.}
    \label{fig:fuse1b}
  \end{figure*}
  \end{landscape}
  }
  
  \onlfig{
  \begin{landscape}
  \addtolength{\textwidth}{6.3cm} 
  \addtolength{\evensidemargin}{-3cm}
  \addtolength{\oddsidemargin}{-3cm}
  \begin{figure*}
   \centering
    \includegraphics[trim=0 0 0 -120,height=20.0cm,angle=0]{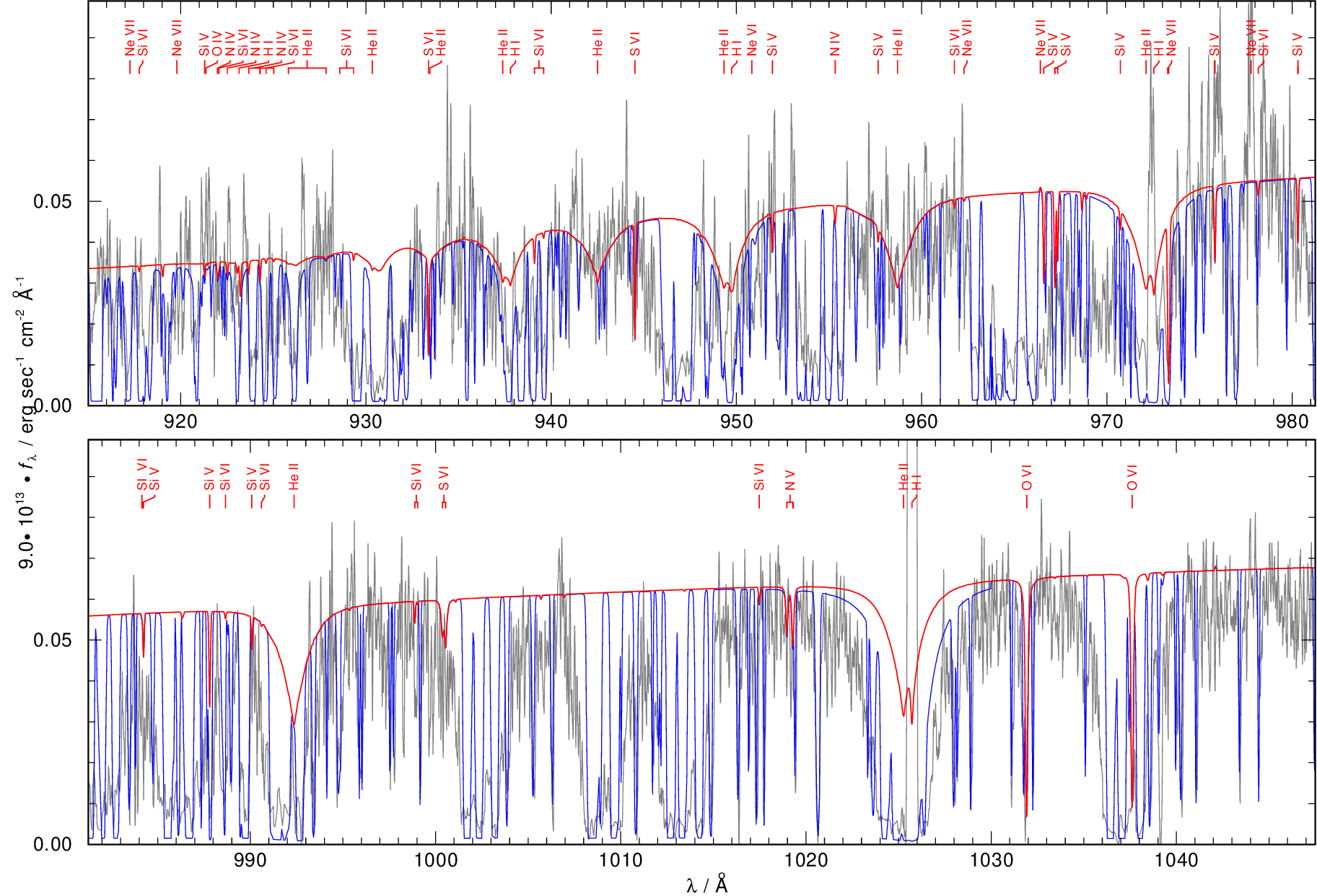}
    \caption{Same as Fig.~\ref{fig:fuse1a} for \pnl.}
    \label{fig:fuse2a}
  \end{figure*}
  \end{landscape}
  }
  
  \addtocounter{figure}{-1}
  
  \onlfig{
  \begin{landscape}
  \addtolength{\textwidth}{6.3cm} 
  \addtolength{\evensidemargin}{-3cm}
  \addtolength{\oddsidemargin}{-3cm}
  \begin{figure*}
   \centering
    \includegraphics[trim=0 0 0 -120,height=20.0cm,angle=0]{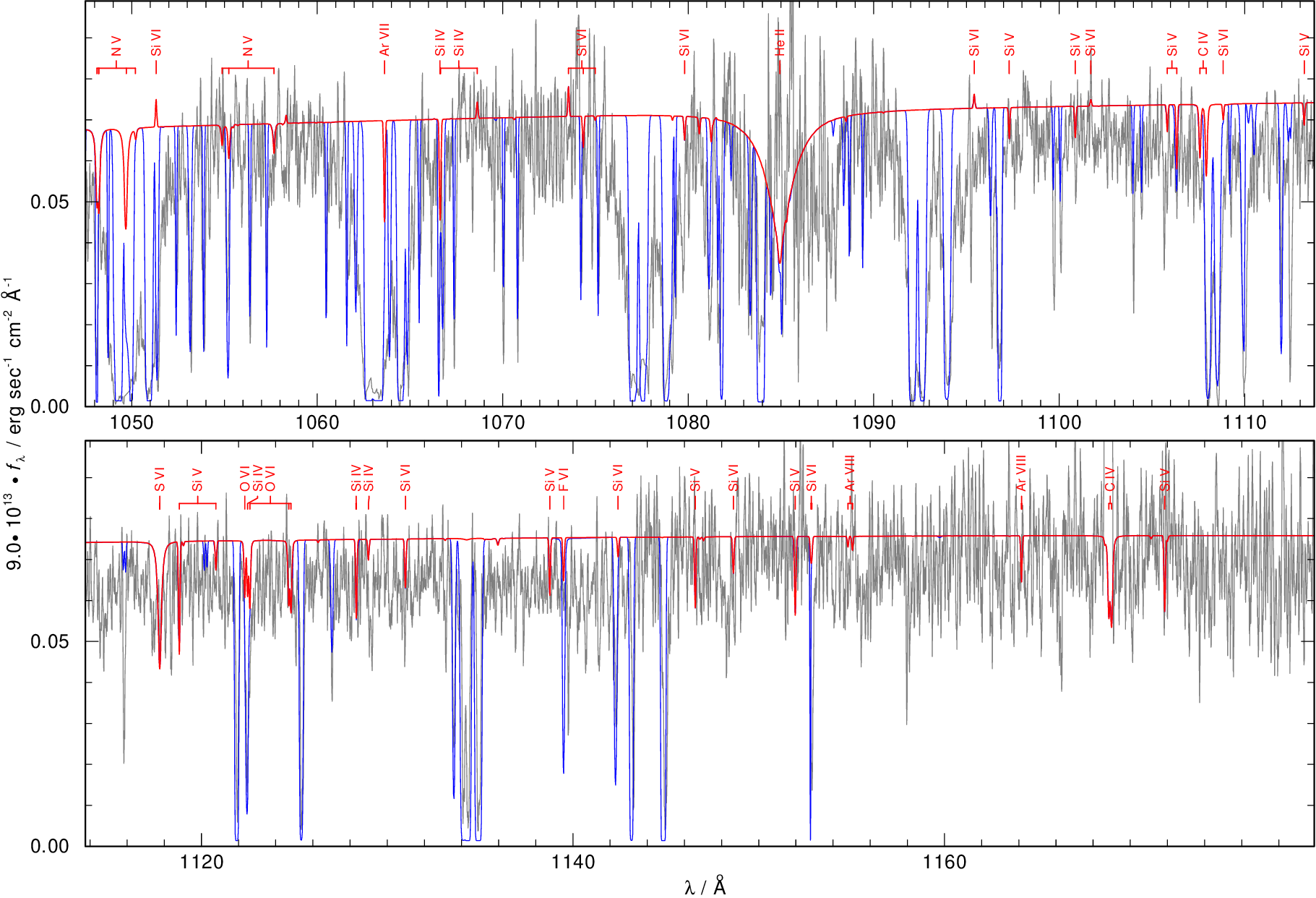}
    \caption{Continued.}
    \label{fig:fuse2b}
  \end{figure*}
  \end{landscape}
  }
  
  \onlfig{
  \begin{landscape}
  \addtolength{\textwidth}{6.3cm} 
  \addtolength{\evensidemargin}{-3cm}
  \addtolength{\oddsidemargin}{-3cm}
  \begin{figure*}
   \centering
    \includegraphics[trim=0 0 0 -120,height=20.0cm,angle=0]{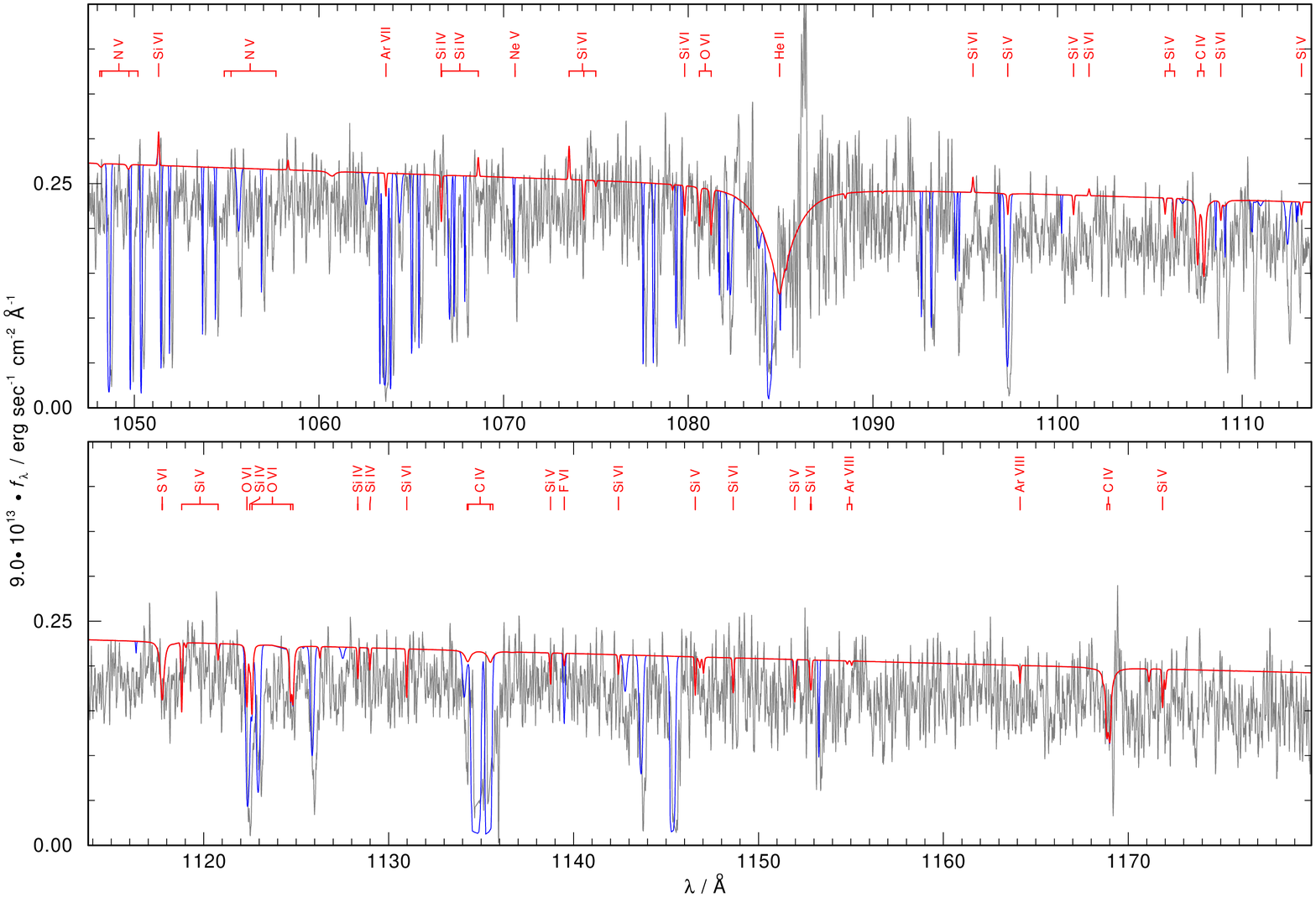}
    \caption{Same as Fig.~\ref{fig:fuse1a} for \hsa.}
    \label{fig:fuse3a}
  \end{figure*}
  \end{landscape}
  }
  
  \addtocounter{figure}{-1}
  
  \onlfig{
  \begin{landscape}
  \addtolength{\textwidth}{6.3cm} 
  \addtolength{\evensidemargin}{-3cm}
  \addtolength{\oddsidemargin}{-3cm}
  \begin{figure*}
   \centering
    \includegraphics[trim=0 0 0 -120,height=20.0cm,angle=0]{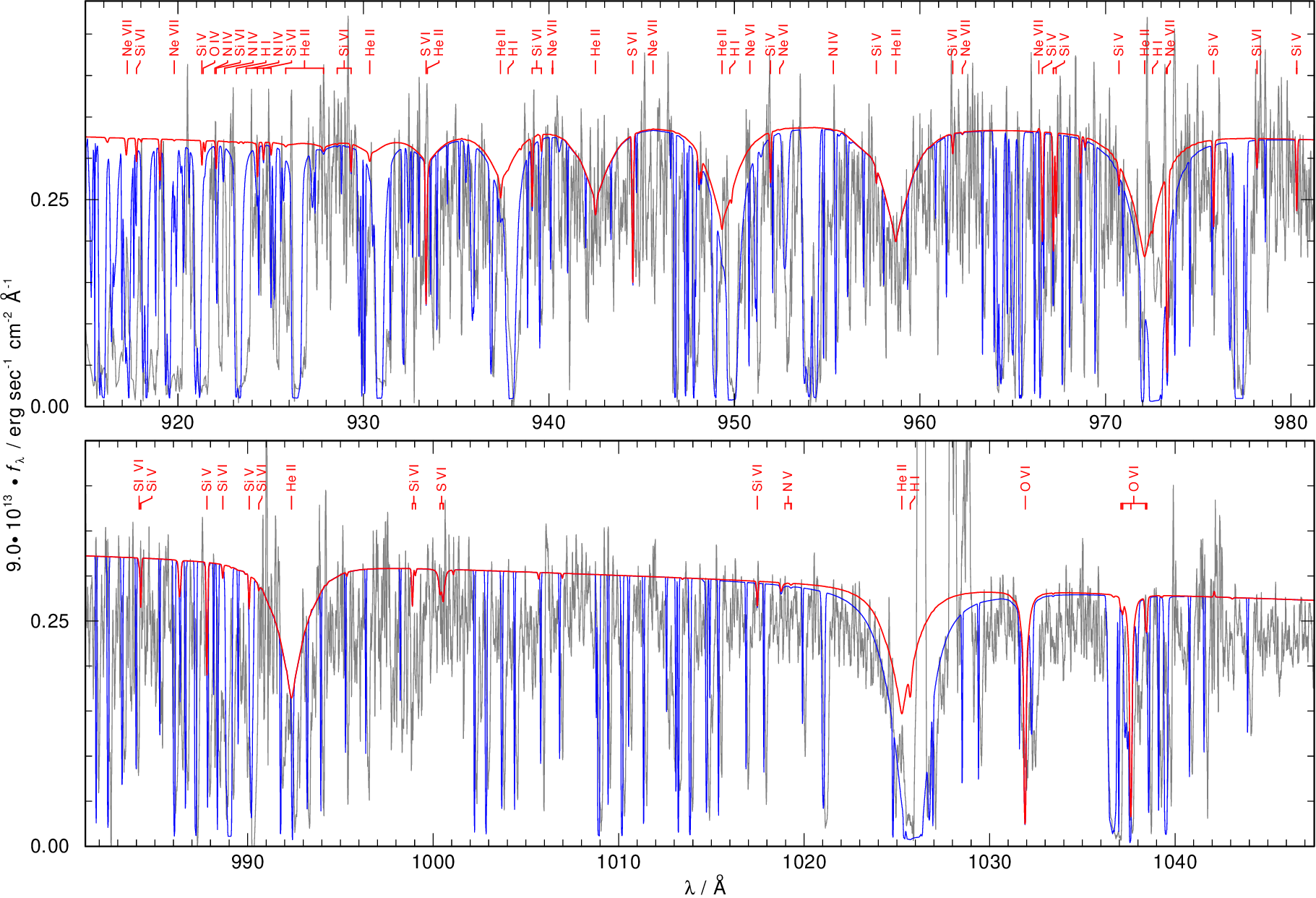}
    \caption{Continued.}
    \label{fig:fuse3b}
  \end{figure*}
  \end{landscape}
  }
  
  \onlfig{
  \begin{landscape}
  \addtolength{\textwidth}{6.3cm} 
  \addtolength{\evensidemargin}{-3cm}
  \addtolength{\oddsidemargin}{-3cm}
  \begin{figure*}
   \centering
    \includegraphics[trim=0 0 0 -120,height=20.0cm,angle=0]{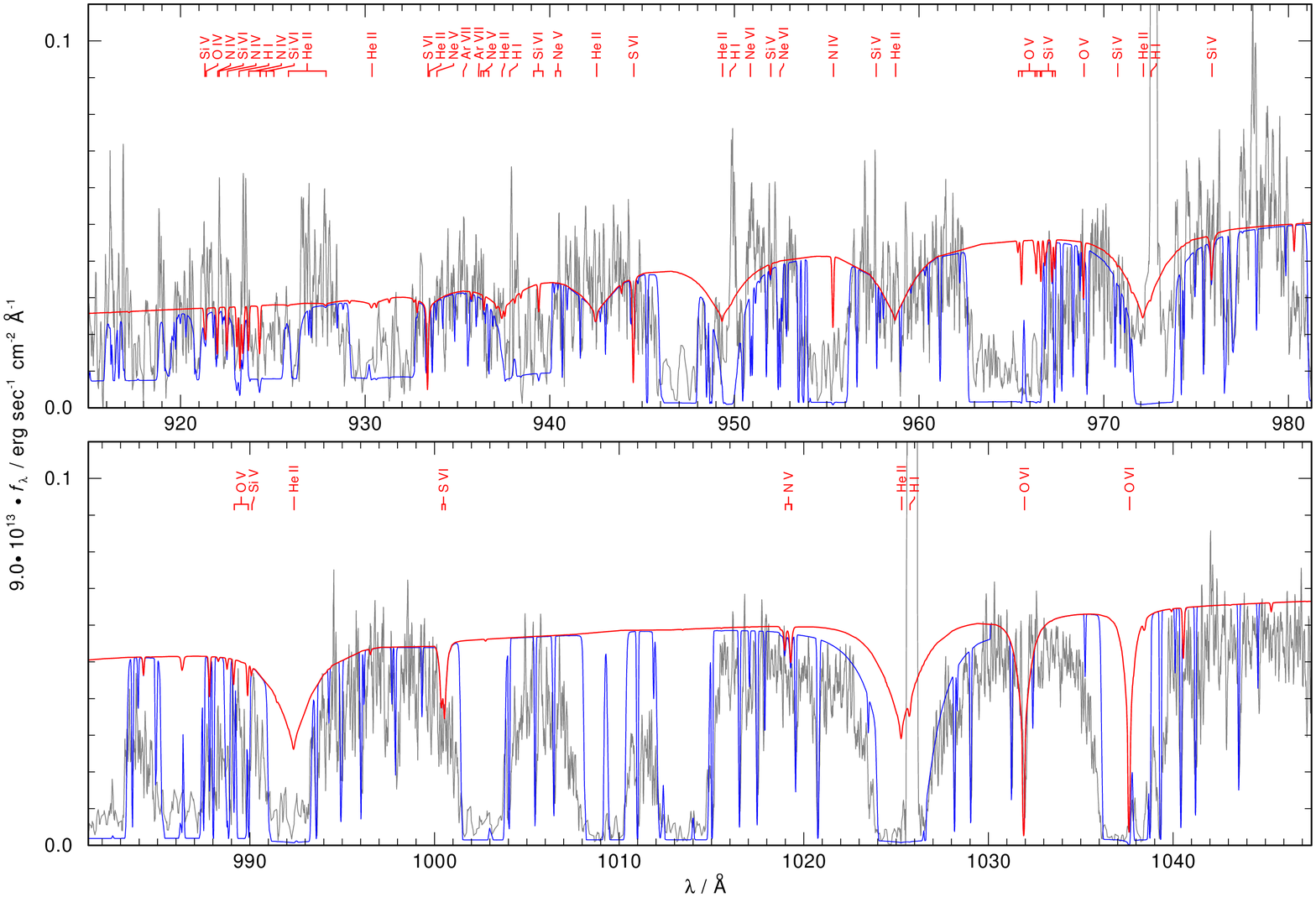}
    \caption{Same as Fig.~\ref{fig:fuse1a} for \hsb.}
    \label{fig:fuse4a}
  \end{figure*}
  \end{landscape}
  }
  
  \addtocounter{figure}{-1}
  
  \onlfig{
  \begin{landscape}
  \addtolength{\textwidth}{6.3cm} 
  \addtolength{\evensidemargin}{-3cm}
  \addtolength{\oddsidemargin}{-3cm}
  \begin{figure*}
   \centering
    \includegraphics[trim=0 0 0 -120,height=20.0cm,angle=0]{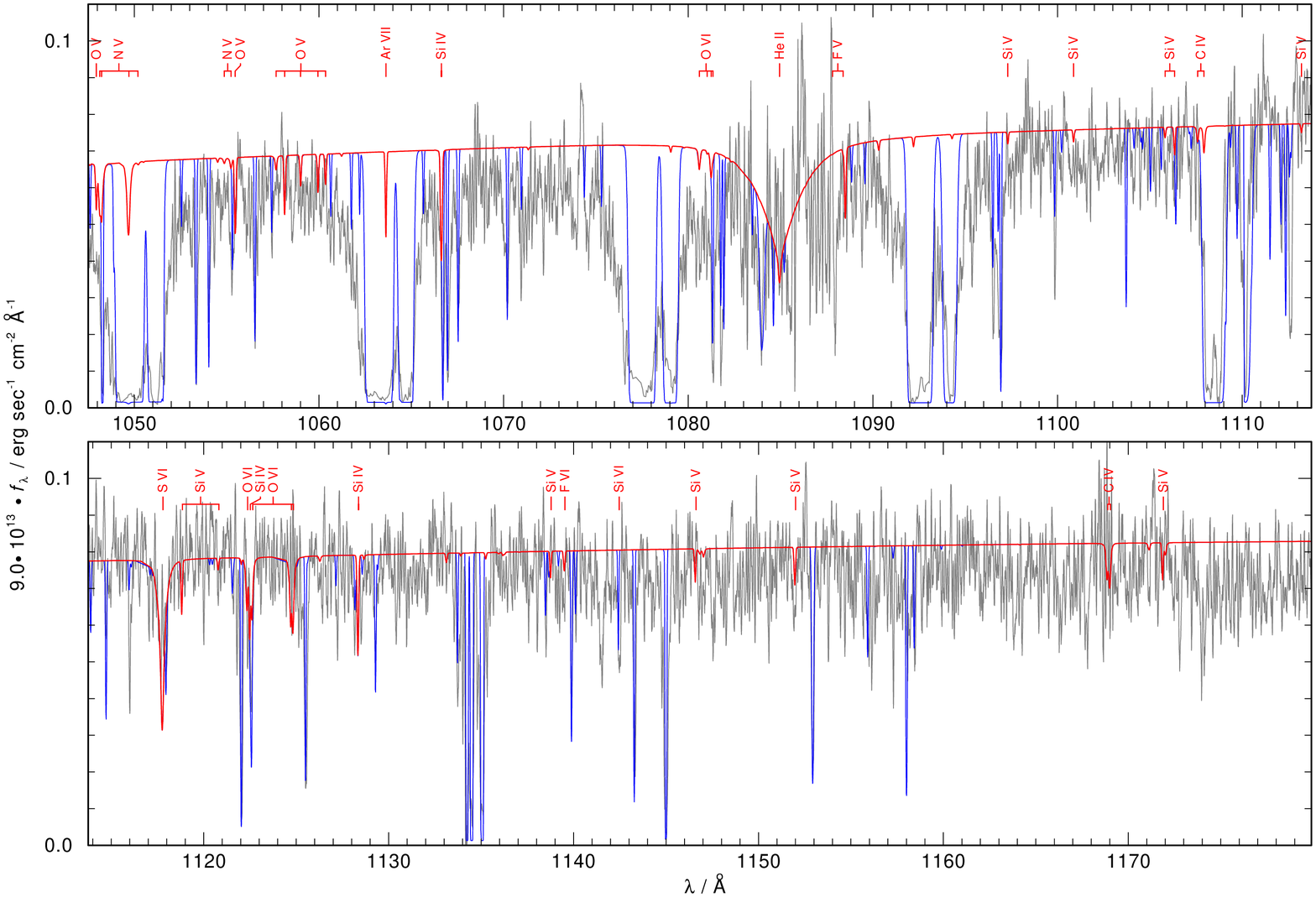}
    \caption{Continued.}
    \label{fig:fuse4b}
  \end{figure*}
  \end{landscape}
  }  

Because of the poor quality of the FUSE spectra, we obtained  
HST/COS\footnote{Cosmic Origins Spectrograph} spectra (Proposal Id: 11699,
Figs.~\ref{fig:cos1} $-$ \ref{fig:cos4}) during April to July 2010 using the grating G140L 
($2000 < R < 3500$ within $1150\,\mathrm{\AA} < \lambda < 1800\,\mathrm{\AA}$)
and the primary science aperture.

\onlfig{
\begin{landscape}
\addtolength{\textwidth}{6.3cm} 
\addtolength{\evensidemargin}{-3cm}
\addtolength{\oddsidemargin}{-3cm}
\begin{figure*}
  \includegraphics[trim=0 0 0 -120,height=20cm,angle=0]{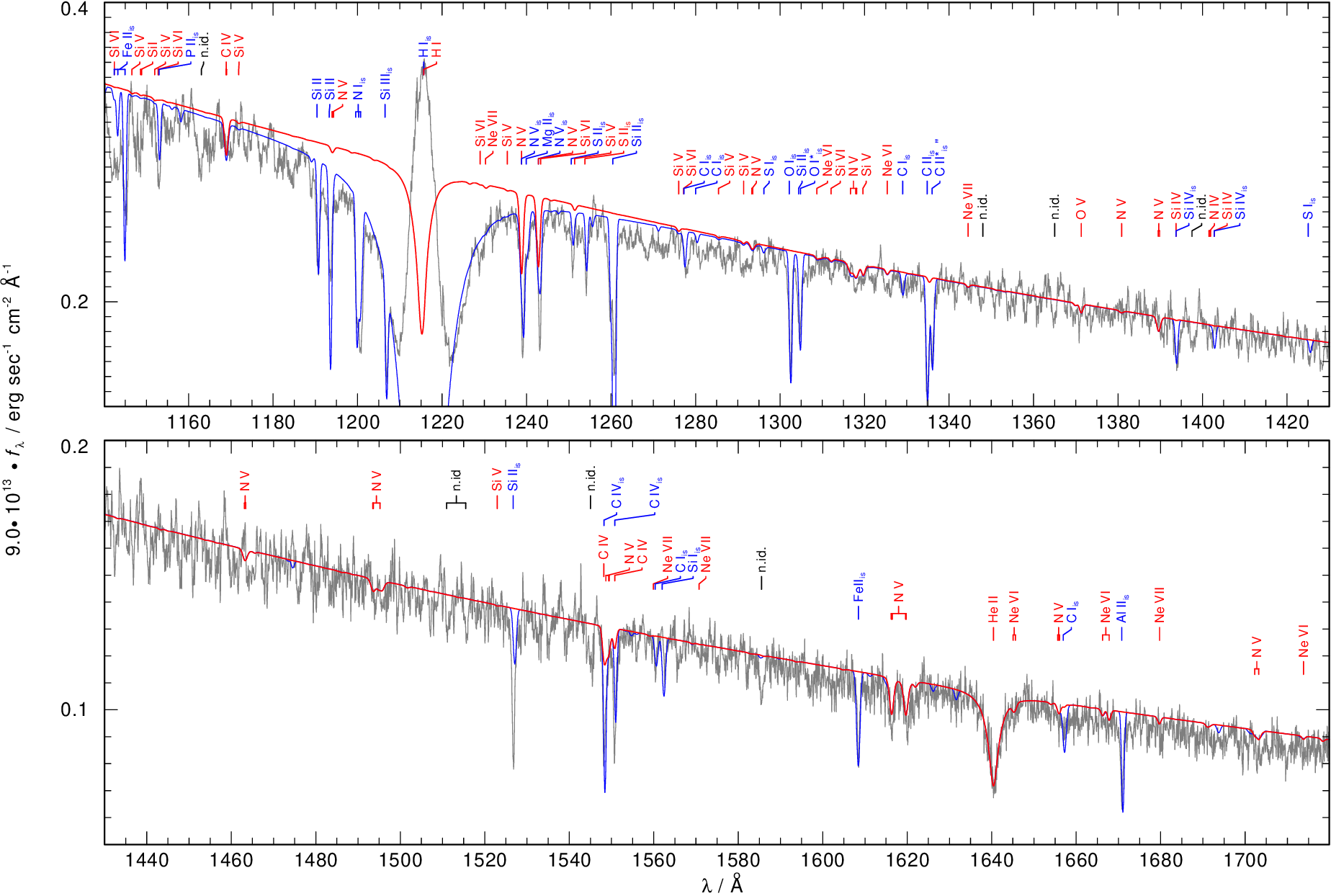}
  \caption{HST/COS spectrum of \pnk (gray) compared with final synthetic spectra (red: pure stellar, 
           blue: combined stellar and interstellar). The locations of photospheric (red) and 
           interstellar (blue) lines reproduced by the synthetic spectrum are marked.}
  \label{fig:cos1}
\end{figure*}
\end{landscape}
}

\onlfig{
\begin{landscape}
\addtolength{\textwidth}{6.3cm} 
\addtolength{\evensidemargin}{-3cm}
\addtolength{\oddsidemargin}{-3cm}
\begin{figure*}
 \centering
  \includegraphics[trim=0 0 0 -120,height=20cm,angle=0]{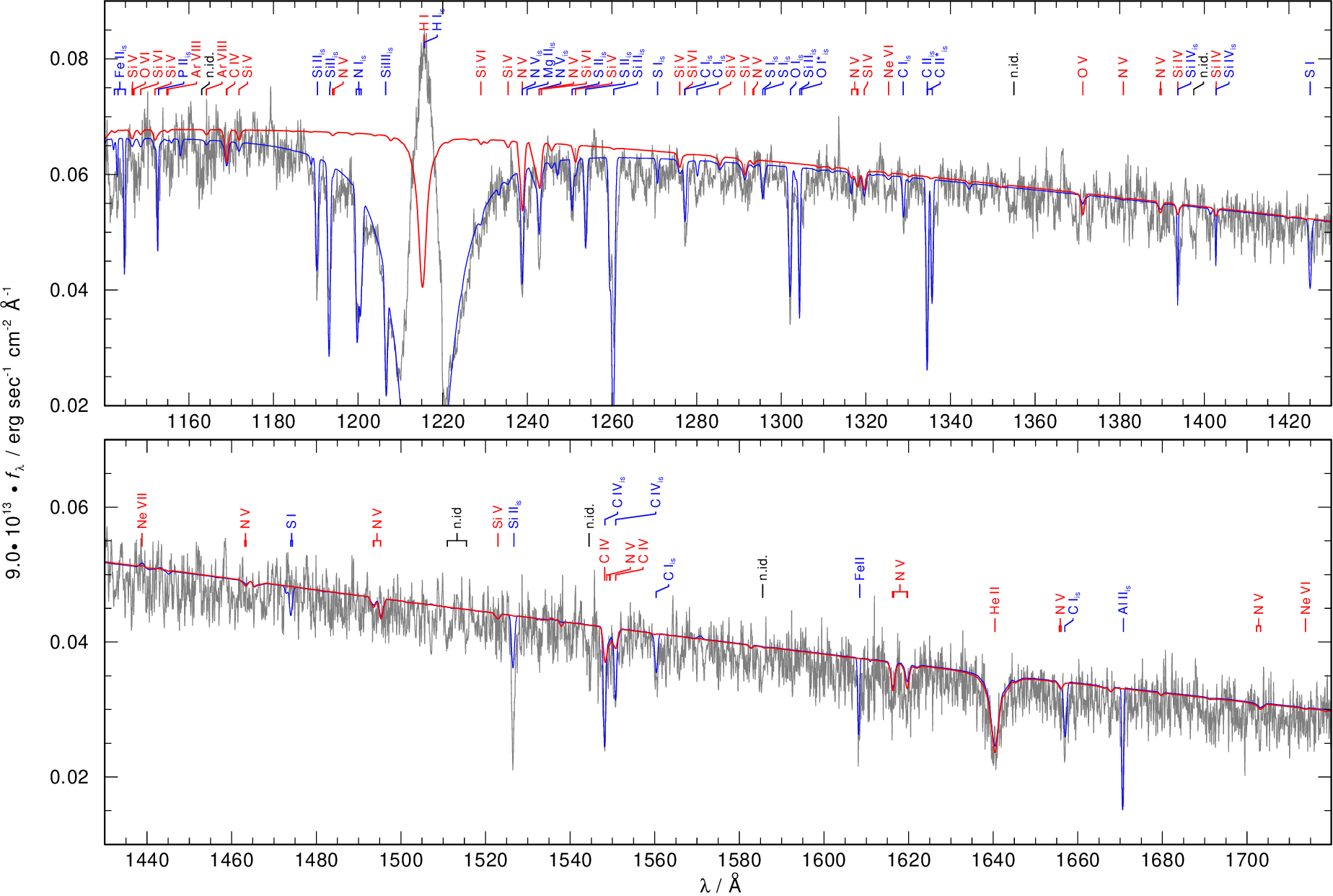}
  \caption{Same as Fig.~\ref{fig:cos1} for \pnl.}
  \label{fig:cos2}
\end{figure*}
\end{landscape}
}

\onlfig{
\begin{landscape}
\addtolength{\textwidth}{6.3cm} 
\addtolength{\evensidemargin}{-3cm}
\addtolength{\oddsidemargin}{-3cm}
\begin{figure*}
 \centering
  \includegraphics[trim=0 0 0 -120,height=20cm,angle=0]{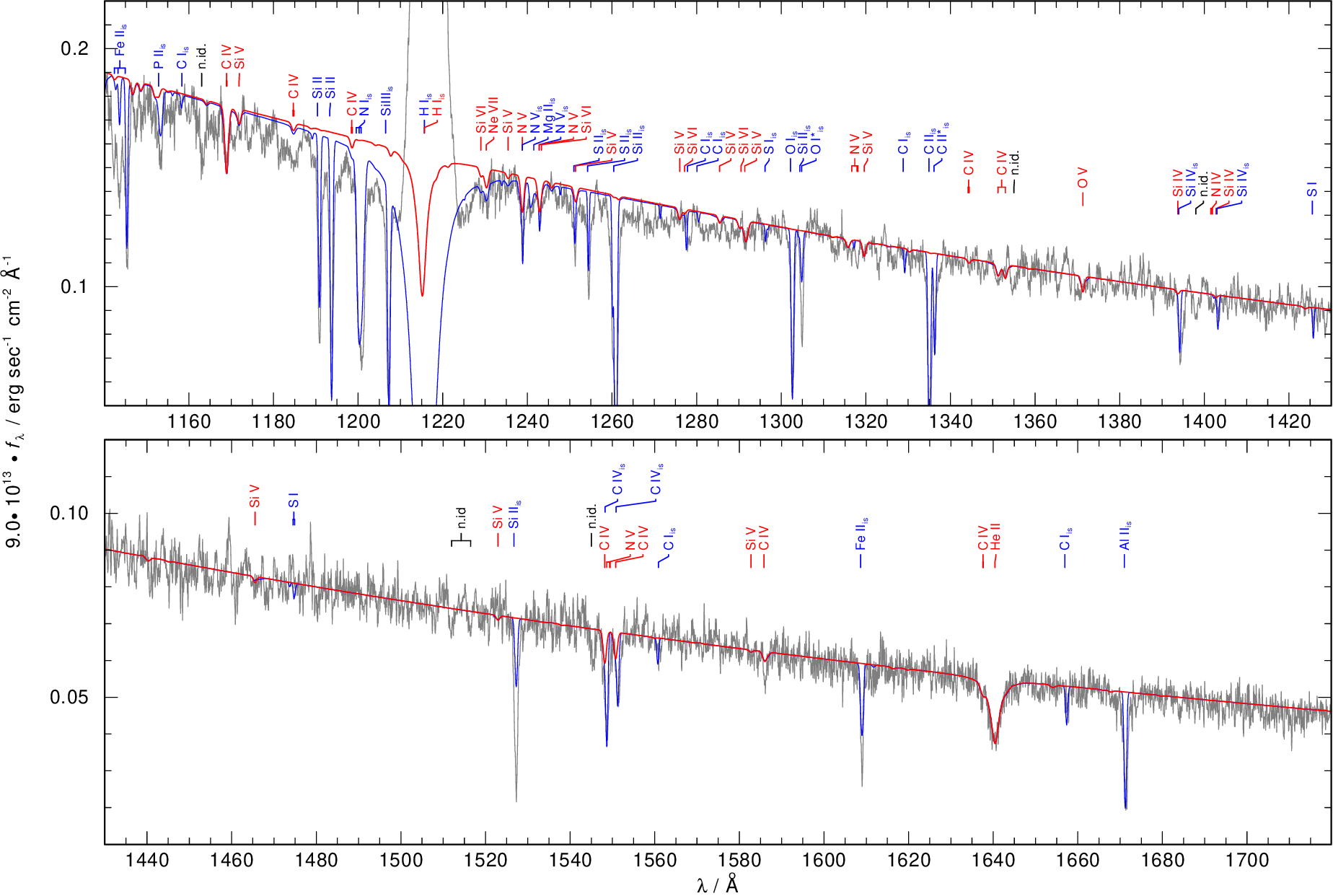}
  \caption{Same as Fig.~\ref{fig:cos1} for \hsa.}
  \label{fig:cos3}
\end{figure*}
\end{landscape}
}

\onlfig{
\begin{landscape}
\addtolength{\textwidth}{6.3cm} 
\addtolength{\evensidemargin}{-3cm}
\addtolength{\oddsidemargin}{-3cm}
\begin{figure*}
 \centering
  \includegraphics[trim=0 0 0 -120,height=20cm,angle=0]{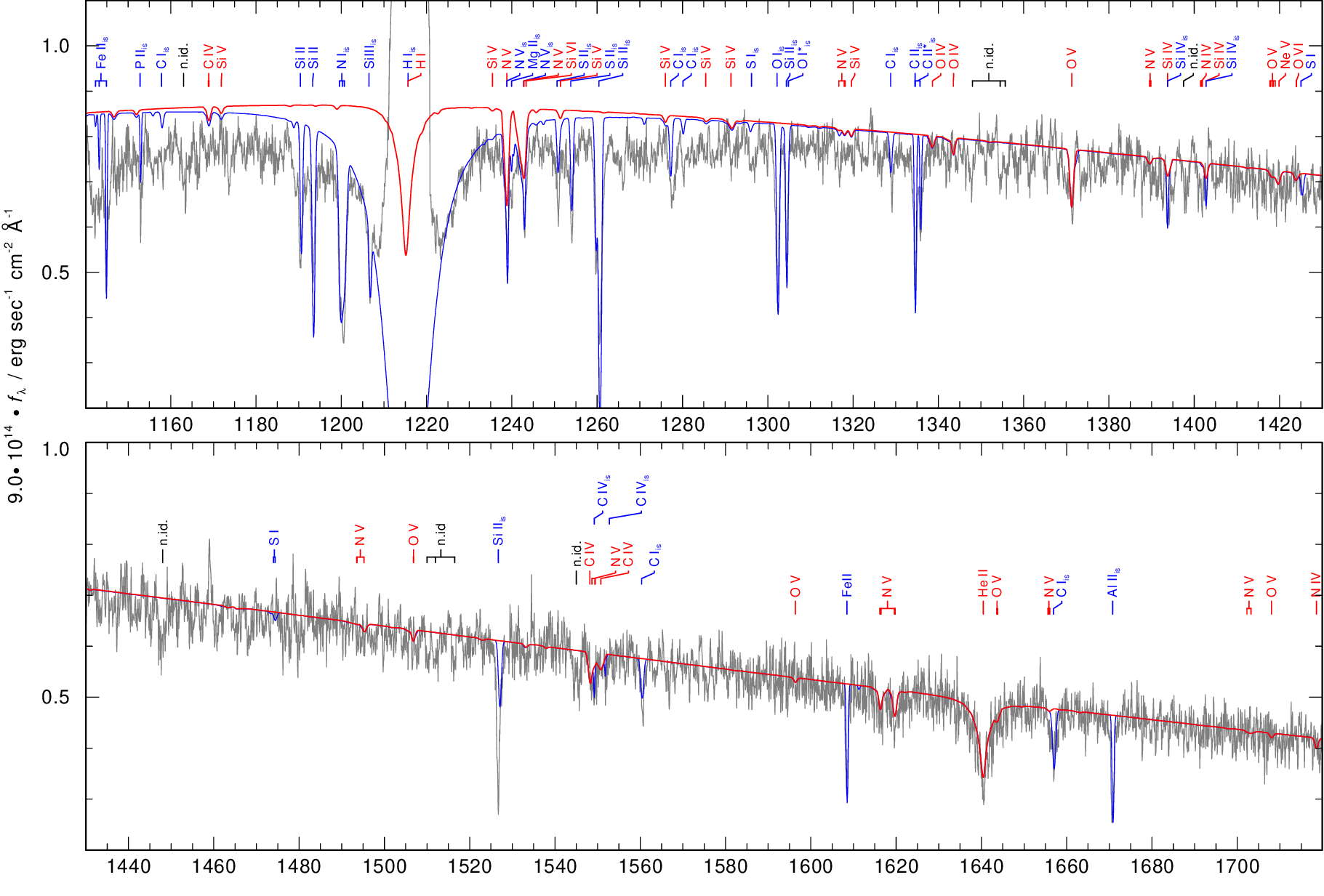}
  \caption{Same as Fig.~\ref{fig:cos1} for \hsb.}
  \label{fig:cos4}
\end{figure*}
\end{landscape}
}

\section{Spectral analysis}
\label{sect:analysis}

Motivated by the new COS spectra, we decided to comprehensively re-analyze all available data of the O(He) stars. 
Since the time when \citet{rauchetal1998} presented their analysis of optical, ultraviolet 
(IUE\footnote{International Ultraviolet Explorer}), and X-ray (ROSAT\footnote{R\"ontgensatellit}) data,
our T{\"u}bingen NLTE Model Atmosphere Package TMAP\footnote{\url{http://astro.uni-tuebingen.de/~TMAP}} 
(Sect.\,\ref{sect:model}) as well as the atomic data, that is taken from 
TMAD\footnote{\url{http://astro.uni-tuebingen.de/~TMAD}}, 
the T\"ubingen Model Atom Database, have continuously improved. 
Moreover, in a standard procedure, we modeled photospheric and interstellar line-absorption spectra to
correctly identify the pure atmospheric lines.
Absorption by interstellar gas was modeled using the program Owens \citep{Hebrardetal2003, HebrardLemoine2002}. 
Owens allows for multiple independent interstellar clouds, each with its own radial and turbulent 
velocities, temperature, and column densities for each of any number of constituent gas species. All properties 
of each cloud are adjusted by fitting Voigt profiles to the data by chi-squared minimization.

\subsection{Model atmospheres}
\label{sect:model}

We used TMAP \citep{werneretal2003,rauchdeetjen2003} to compute plane-parallel, line-blanketed non-LTE model 
atmospheres in radiative and hydrostatic equilibrium. In the previous analysis of the O(He) stars, 
\citet{rauchetal1994, rauchetal1996, rauchetal1998} used only H+He model atmospheres to derive \Teff, \logg, and 
the H/He ratio. In our analysis, we also included the elements C, N, O, F, Ne, Si, P, S, and Ar with their dominant 
ionization stages (Table ~\ref{tab:statistics}, Figs.~\ref{fig:ionk} $-$ \ref{fig:ionhb}) to study their impact on 
the model atmospheres and the resulting line profiles. 
We found that C, N, O, and Ne have a strong influence on the atmospheric structure. 
In Fig.~\ref{fig:Ne} we demonstrate the strong effect of additional opacities on the resulting line profiles. 
The impact of F, Si, P, S, and Ar turned out to be negligible. \citet{DreizlerWerner1994} have shown that line 
blanketing by iron-group elements hardly affects the hydrogen and helium lines in hot central stars. To 
compute a model grid in reasonable time, we therefore decided to include only H, He, C, N, O, and Ne in our model 
atmospheres to derive \Teff, \logg, and the element abundances of these elements. To determine the 
abundances of F, Si, P, S, Ar, and Fe, we kept the values of \Teff\ and \logg\ fixed. 
The upper limits (Table\,\ref{tab:parameters}) were derived by test models where the respective lines in the model 
contradict the non-detection of the lines in the observation (at the abundance limit).

\onlfig{
\begin{figure*}
  \resizebox{0.7\hsize}{!}{\includegraphics{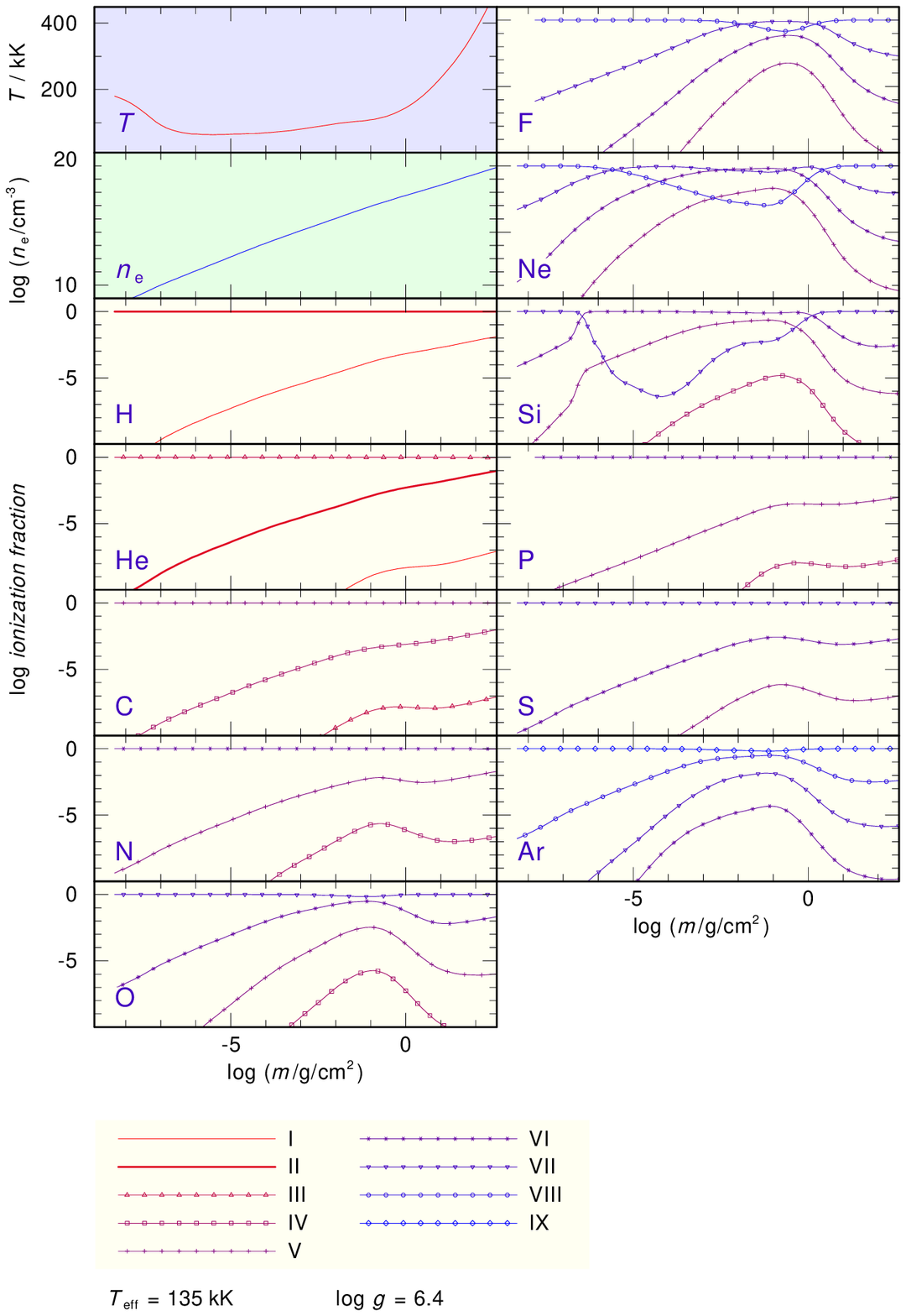}}
  \caption{Temperature, electron density stratification, 
           and ionization fractions of all elements in our final model of \pnk.} 
  \label{fig:ionk}
\end{figure*}
}

\onlfig{
\begin{figure*}
  \resizebox{0.7\hsize}{!}{\includegraphics{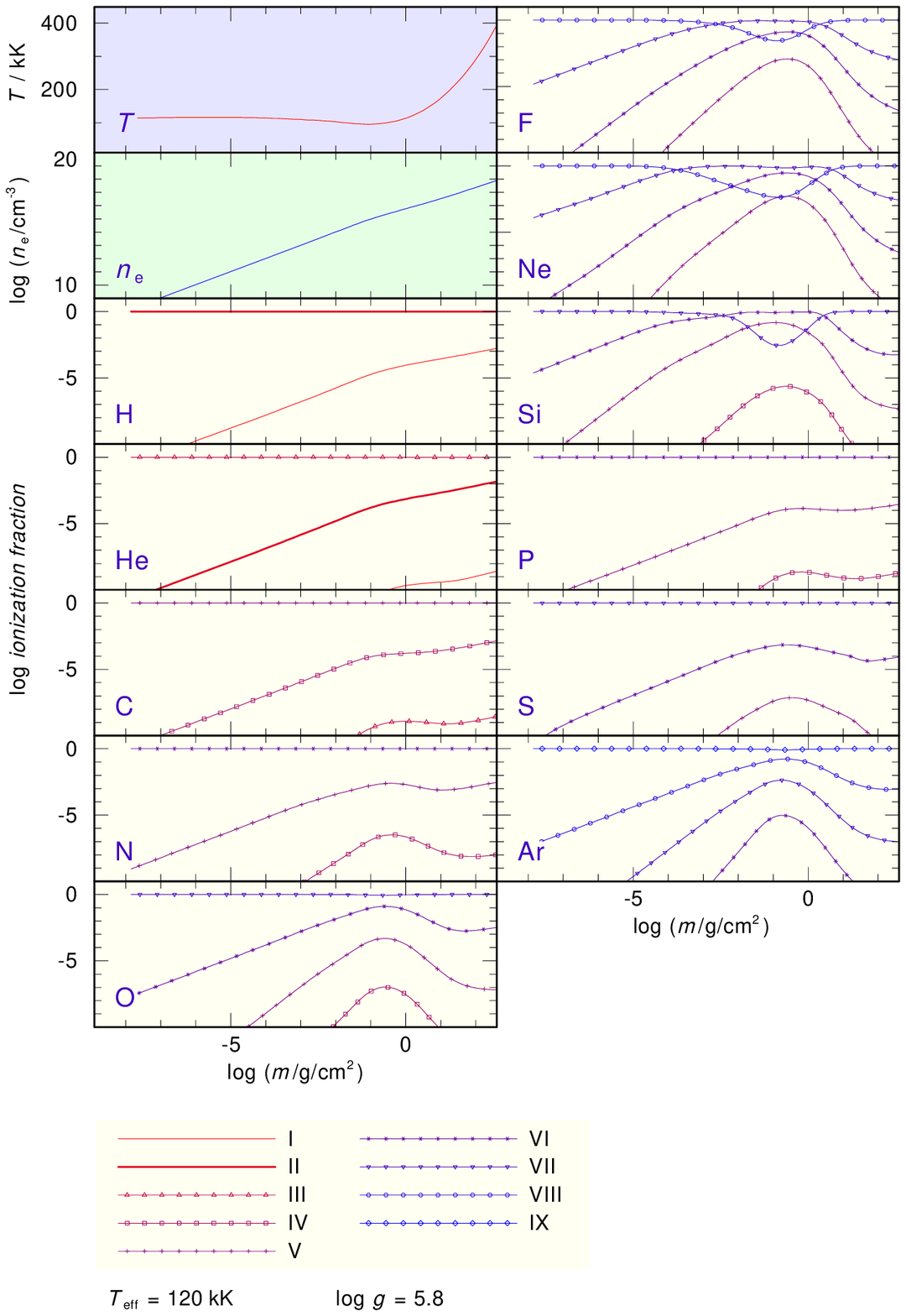}}
  \caption{Like Fig.~\ref{fig:ionk}, for \pnl.} 
  \label{fig:ionl}
\end{figure*}
}

\onlfig{
\begin{figure*}
  \resizebox{0.7\hsize}{!}{\includegraphics{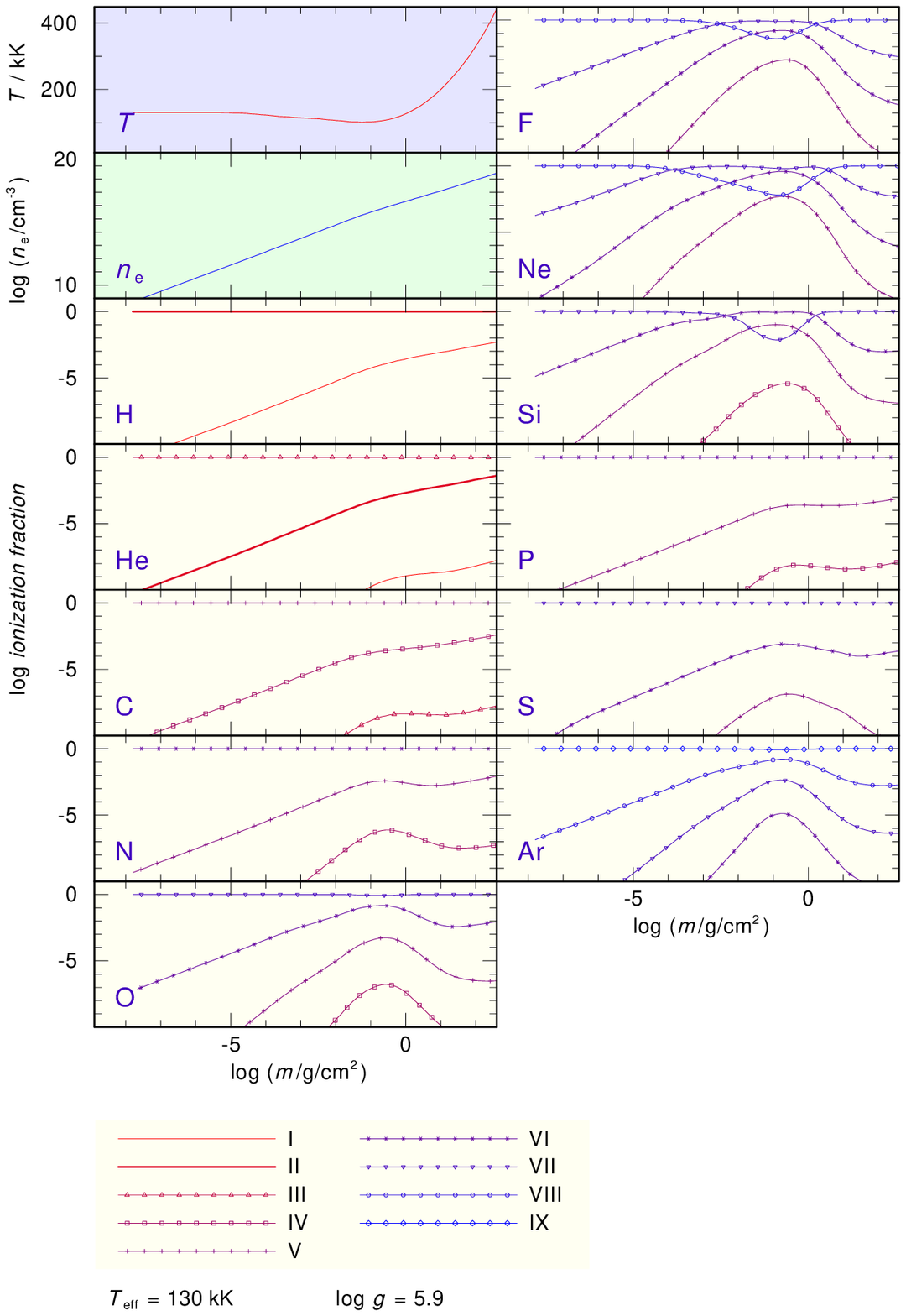}}
  \caption{Like Fig.~\ref{fig:ionk}, for \hsa.} 
  \label{fig:ionkha}
\end{figure*}
}

\onlfig{
\begin{figure*}
  \resizebox{0.7\hsize}{!}{\includegraphics{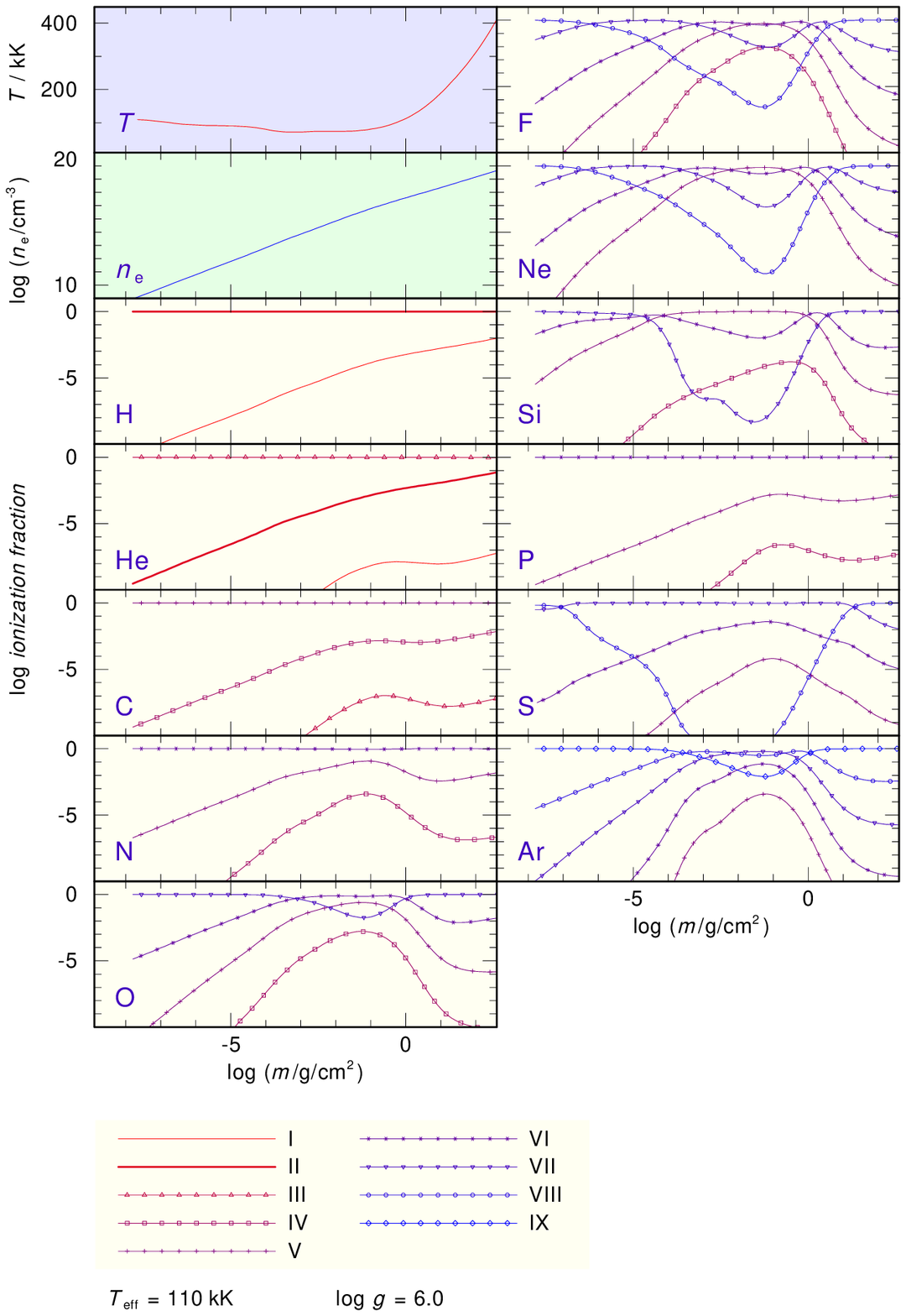}}
  \caption{Like Fig.~\ref{fig:ionk}, for \hsb.} 
  \label{fig:ionhb}
\end{figure*}
}

\onltab{
\begin{table}
\centering
\caption{Statistics of the model atoms used in our TMAP calculations.}
\label{tab:statistics}
\begin{tabular}{rccccccc}
\hline
\hline
               & \multicolumn{7}{c}{\Teff} \\
\cline{2-8}
\noalign{\smallskip}                                                   
               & \multicolumn{3}{c}{$>110$\,kK} && \multicolumn{3}{c}{$\leq 110$\,kK} \\
\cline{2-4}
\cline{6-8}
\noalign{\smallskip}                                                   
ion            & NLTE & LTE & lines && NLTE & LTE & lines \\
 \hline         
\noalign{\smallskip}                                                   
H\,{\sc i}     &   10 &   6 &    45 &&   10 &   6 &    45 \\
H\,{\sc ii}    &    1 & $-$ &   $-$ &&    1 & $-$ &   $-$ \\
\noalign{\smallskip}                                                   
He\,{\sc i}    &    5 &  98 &     3 &&    5 &  98 &     3 \\
He\,{\sc ii}   &   14 &  18 &    91 &&   14 &  18 &    91 \\
He\,{\sc iii}  &    1 & $-$ &   $-$ &&    1 & $-$ &   $-$ \\
\noalign{\smallskip}                                                   
C\,{\sc iii}   &   13 &  54 &    32 &&   13 &  54 &    32 \\
C\,{\sc iv}    &   14 &  44 &    35 &&   14 &  44 &    35 \\
C\,{\sc v}     &    1 &   0 &     0 &&    1 &   0 &     0 \\
\noalign{\smallskip}                                                   
N\,{\sc iii}   &    1 &  65 &     0 &&    1 &  65 &     0 \\
N\,{\sc iv}    &   16 &  78 &    30 &&   16 &  78 &    30 \\
N\,{\sc v}     &   14 &  48 &    35 &&   14 &  48 &    35 \\
N\,{\sc vi}    &    1 &   0 &     0 &&    1 &   0 &     0 \\    
\noalign{\smallskip}                                                  
O\,{\sc iv}    &   18 &  76 &    39 &&   18 &  76 &    39 \\
O\,{\sc v}     &   17 & 109 &    35 &&   17 & 109 &    35 \\
O\,{\sc vi}    &   14 &  48 &    33 &&   14 &  48 &    33 \\
O\,{\sc vii}   &    1 &   0 &     0 &&    1 &   0 &     0 \\    
\noalign{\smallskip}                                                   
F\,{\sc v}     &    1 &  10 &     0 &&    1 &  10 &     0 \\
F\,{\sc vi}    &    6 &   6 &     0 &&    6 &   6 &     0 \\
F\,{\sc vii}   &    2 &   4 &     0 &&    2 &   4 &     0 \\
F\,{\sc viii}  &    1 &   0 &     0 &&    1 &   0 &     0 \\    
\noalign{\smallskip}                                                   
Ne\,{\sc iv}   &    0 &   0 &     0 &&    2 &  39 &     0 \\
Ne\,{\sc v}    &   14 &  80 &    18 &&   14 &  80 &    18 \\
Ne\,{\sc vi}   &   14 &  17 &    30 &&   14 &  17 &    30 \\
Ne\,{\sc vii}  &   15 &  94 &    27 &&   15 &  94 &    27 \\
Ne\,{\sc viii} &   14 &  90 &    35 &&    1 &   0 &     0 \\
Ne\,{\sc   ix} &    1 &   0 &     0 &&    1 &   0 &     0 \\
\noalign{\smallskip}                                                   
Si\,{\sc iv}   &   12 &  11 &    24 &&   12 &  11 &    24 \\
Si\,{\sc v}    &   25 &   0 &    59 &&   25 &   0 &    59 \\
Si\,{\sc vi}   &   45 & 195 &   193 &&   45 & 195 &   193 \\
Si\,{\sc vii}  &    1 &   0 &     0 &&    1 &   0 &     0 \\
\noalign{\smallskip}                                                   
P\,{\sc iv}    &   15 &  36 &     9 &&   15 &  36 &     9 \\
P\,{\sc v}     &   18 &   7 &    12 &&   18 &   7 &    12 \\
P\,{\sc vi}    &    1 &   0 &     0 &&    1 &   0 &     0 \\
\noalign{\smallskip}                                                   
S\,{\sc v}     &   23 &  87 &    47 &&   23 &  87 &    47 \\
S\,{\sc vi}    &   25 &  12 &    25 &&   25 &  12 &    25 \\
S\,{\sc vii}   &    1 &   0 &     0 &&    1 &   0 &     0 \\
\noalign{\smallskip}                                               
Ar\,{\sc v}    &      &     &       &&    1 & 359 &     0 \\
Ar\,{\sc vi}   &    1 & 183 &     0 &&   14 & 170 &    16 \\
Ar\,{\sc vii}  &   40 & 112 &   130 &&   40 & 112 &   130 \\
Ar\,{\sc viii} &   13 &  28 &    24 &&   13 &  28 &    24 \\
Ar\,{\sc ix}   &    1 &   0 &     0 &&    1 &   0 &     0 \\
\hline 
\end{tabular}
\end{table}
}

\begin{figure}
  \resizebox{\hsize}{!}{\includegraphics{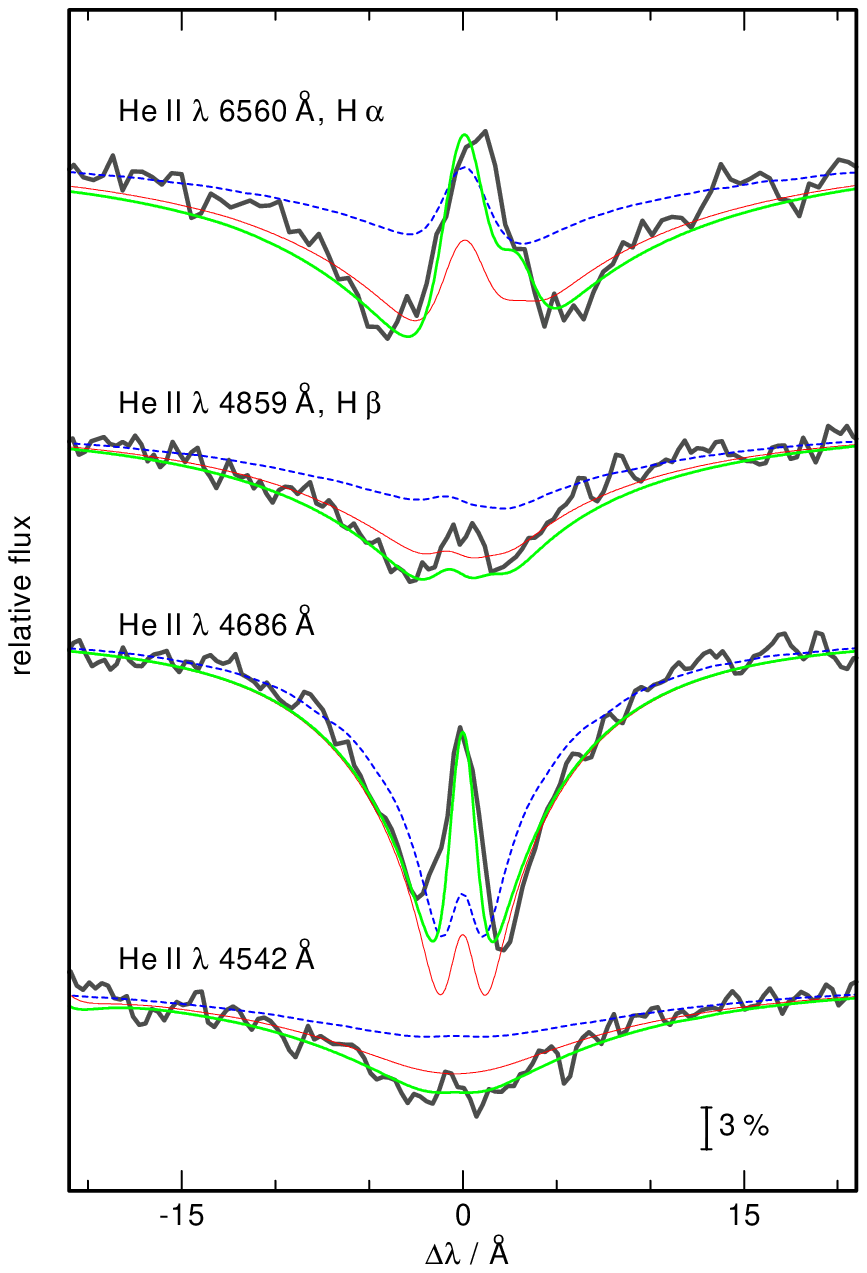}}
  \caption{Comparison of optical \Ion{H}{I} and \Ion{He}{II} lines calculated from models including H+He (dashed, blue), 
    H+He+C+N+O (thin, red), and H+He+C+N+O+Ne (thick, green) with \Teffw{135} and \loggw{6.4}, overplotted on the EFOSC2 spectrum 
    of \pnk. The vertical bar indicates 3\,\% of the continuum flux.}
  \label{fig:Ne}
\end{figure}

\subsection{Interstellar neutral hydrogen and reddening}
\label{sect:ism}

\begin{figure}
  \resizebox{\hsize}{!}{\includegraphics{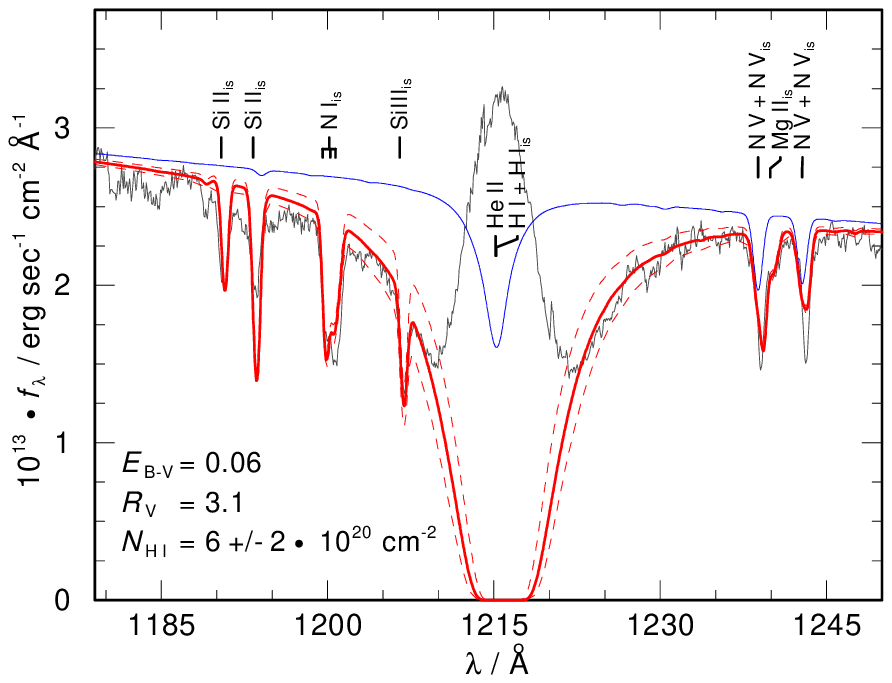}}
  \caption{Determination of the interstellar \Ion{H}{I} column density toward \pnk.
           The dashed lines indicate the error limits.
           The thin (blue) line is the pure photospheric spectrum.
           Prominent spectral lines are marked at top, 
           ``is'' indicates interstellar origin.
           The line center of Ly\,$\alpha$ is dominated by
           geocoronal emission.
           }
  \label{fig:NHI}
\end{figure}

\begin{figure}
  \resizebox{\hsize}{!}{\includegraphics{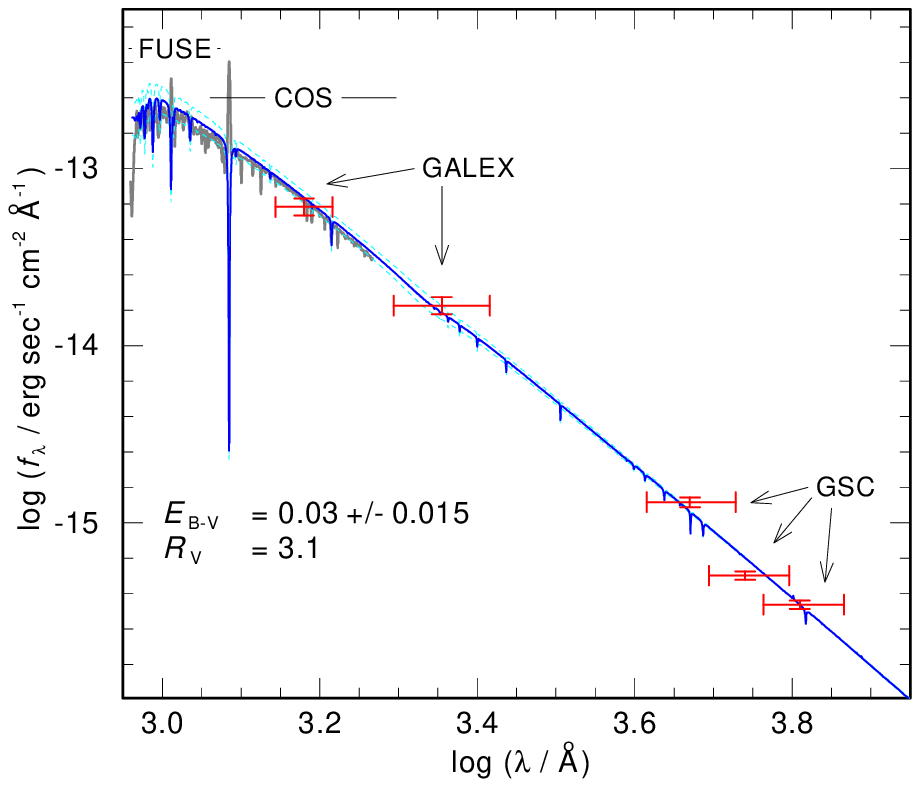}}
  \caption{Determination of \ebv for \hsa.
           The FUSE and COS spectra and GALEX and BVR brightnesses from the GSC\,2.3 catalog are used for comparison.
           The thick (blue) line is the spectrum of the final photospheric model. The dashed lines indicate the error limits.} 
  \label{fig:EBV}
\end{figure}

The interstellar \Ion{H}{I} column density was measured from Ly\,$\alpha$ (Fig.~\ref{fig:NHI}). 
Interstellar reddening was determined using the reddening law of \citet{fitzpatrick1999}.
Since its impact is negligible in the infrared, the model flux was normalized to the most reddest 
brightness found in the literature. Fig.~\ref{fig:EBV} shows the example of \hsa, where the
GSC\footnote{Guide Star Catalogue, \citet{Bucciarelli2001}} R brightness is used. Table~\ref{tab:EBVNHI} summarizes $N_\mathrm{H\,I}$ 
and \ebv of the O(He) stars compared with GALEX\footnote{Galaxy Evolution Explorer, \url{http://www.galex.caltech.edu/}} values, 
which agree well. Within the error limits, our values agree with those of \citet{rauchetal1998}. 

\begin{table}
\caption{Interstellar \Ion{H}{I} column density and reddening.}
  \label{tab:EBVNHI}
\centering                                     
\begin{tabular}{cr@{\,$\pm$\,}lr@{\,$\pm$\,}lc}       
\hline
\hline                    
            & \multicolumn{2}{c}{}                                             & \multicolumn{3}{c}{$E_{B-V}$}         \\
\noalign{\smallskip}
\cline{4-6}
\noalign{\smallskip}
object      & \multicolumn{2}{c}{$N_\mathrm{H\,I}\,[10^{20}\mathrm{cm}^{-2}]$} & \multicolumn{2}{c}{this work} & GALEX \\
\noalign{\smallskip}
\hline
\noalign{\smallskip}
\pnk        &                     6.0 & 2.0 & 0.06 & 0.020 & 0.0797  \\
\pnl        & \hbox{}\hspace{3mm}10.0 & 1.5 & 0.20 & 0.020 & 0.1893  \\
\hsa        &                     2.0 & 1.0 & 0.03 & 0.015 & 0.0250  \\ 
\hsb        &                     5.0 & 1.0 & 0.23 & 0.040 &         \\
\hline
\end{tabular}
\end{table}

\subsection{Effective temperature and surface gravity}
\label{sect:tefflogg}

\begin{figure}
  \resizebox{\hsize}{!}{\includegraphics{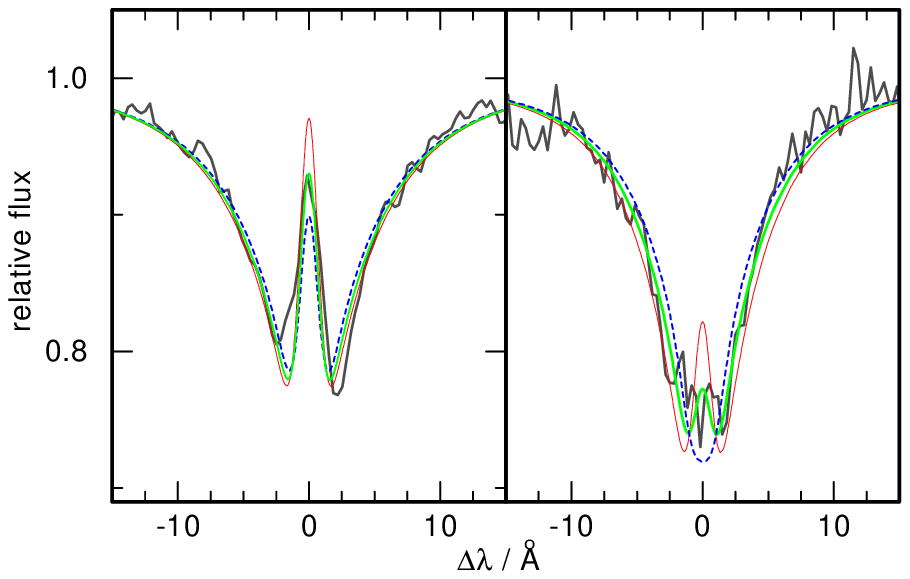}}
  \caption{Determination of \Teff\ of \pnk (left) and \pnl (right) using the central emission of \Ionw{He}{II}{4686}. 
    The EFOSC2 observation (solid black line) is compared 
           with models with \Teff\ = 
           130/110\,kK  (dashed, blue),
           135/120\,kK  (thick, green), and 140/130\,kK (thin, red).}
  \label{fig:TeffKL}
\end{figure}

To determine of \Teff\ and \logg, we used the optical \Ion{H}{I} and \Ion{He}{II} lines. 
For \hsa and \hsb we additionally used the O ionization equilibria to constrain \Teff. 
The new optical observation of \pnk and \pnl allowed us to use the central emission of 
\Ionw{He}{II}{4686} to derive \Teff\ for the two CSPNe (Fig.~\ref{fig:TeffKL}).

For \pnk, \Teffw{135} and \loggw{6.4} (cm/s) were found. The value of \logg\ agrees 
with \citet[\loggw{6.5\pm0.5}]{rauchetal1994,rauchetal1998}, but for \Teff\ we found 
a large deviation from the result of \citet[\Teffw{105\pm10}]{rauchetal1998}.  
However, \citet{rauchetal1994,rauchetal1998} encountered problems with the lower value of \Teff\, 
which can be solved with \Teffw{135}: 
first, with \Teffw{105} the star did not provide enough hard photons to ionize its nebula. With the 
higher \Teff\ for \pnk, we achieve a more consistent PN $\leftrightarrow$ CSPN model. 
Second, \Ionww{N}{V}{4604, 4620} do not appear in emission at \Teffw{105} (as the observation shows), 
but for \Teffw{135} they do. 
Finally, \citet{rauchetal1994} had the problem that the central depression of the \Ionw{He}{II}{4686} 
(n $-$ n' = 3 $-$ 4) line was too strong, while other \Ion{H}{I} and \Ion{He}{II} lines were almost 
perfectly reproduced. The addition of Ne to the models solves this problem. Furthermore, the central 
\Ionw{He}{II}{4686} and H\,$\alpha$ emission can only be reproduced with a supersolar Ne abundance, which causes 
a strong decrease of the temperature in the outer atmosphere. \Ionw{He}{II}{4686} and H\,$\alpha$ are 
most affected by this because they are formed at lower Rosseland optical depths. 
We stress that the large deviation from the previous \Teff\ value is a result of the additional opacities used 
in our model atmospheres, as mentioned above. 

\begin{figure}
  \resizebox{\hsize}{!}{\includegraphics{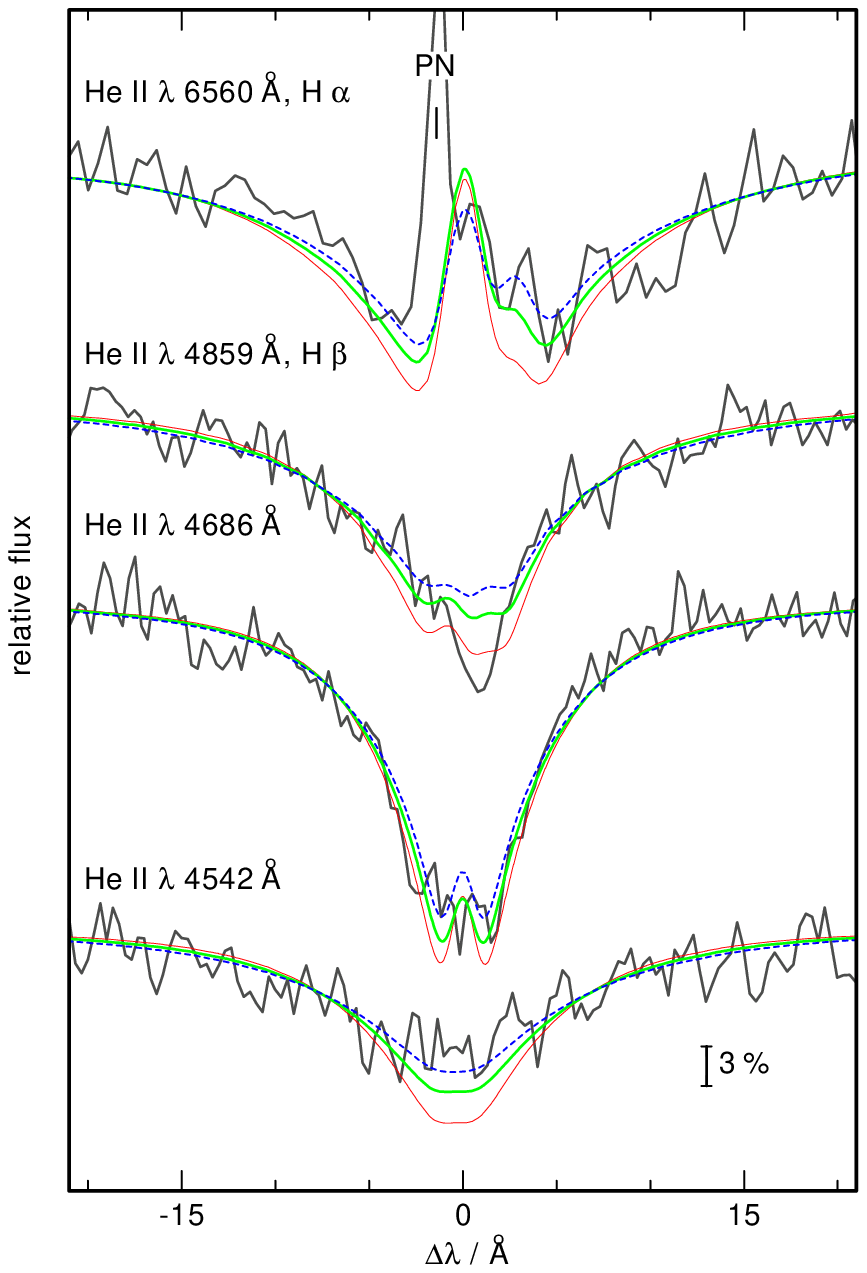}}
  \caption{Determination of \logg\ of \pnl. The EFOSC2 observation (solid black line) is compared 
           with models with \logg\ = 
           6.0 (dashed, blue),
            5.8 (thick, green), and
           5.6 (thin, red). The vertical bar indicates 3\,\% of the continuum flux.}
  \label{fig:logg}
\end{figure}

For \pnl, we confirmed the value of \Teffw{120} \citep{rauchetal1996} but increased the lower limit to 
\Teffw{115} because for lower values, \Ionww{N}{V}{4604, 4620} do not appear in emission as in the observation. 
The new optical spectra allowed us a better determination of the surface gravity (Fig.~\ref{fig:logg}). We found 
\loggw{5.8 \pm 0.2}, which is within the error limits of the literature value \citep[\loggw{5.5\pm0.5}]{rauchetal1996}.

For \hsa, we determined a lower value of \Teffw{130} at a higher \loggw{5.9}
\citep[compared with ][\Teffw{140\pm10}, \loggw{5.5\pm0.5}]{rauchetal1998}. 
The errors of \logg\ were reduced to $\pm$\,0.2.

\begin{figure}
  \resizebox{\hsize}{!}{\includegraphics{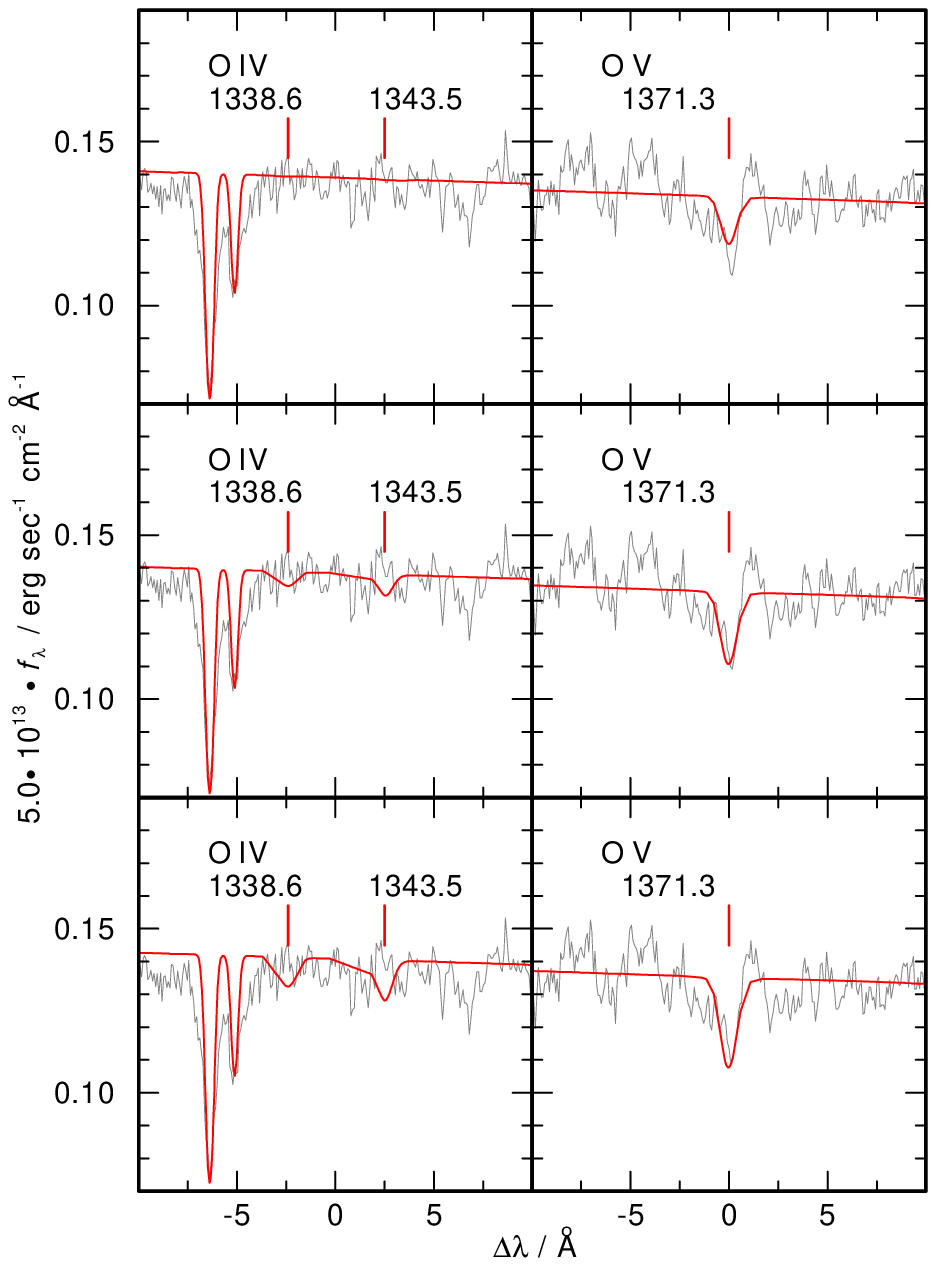}}
  \caption{Determination of \Teff\ of \hsb from the \Ion{O}{IV}\,/\,\Ion{O}{V} ionization equilibrium.
           Top panels: \Teffw{100}, middle: \Teffw{110}; bottom: \Teffw{120}.}
  \label{fig:TeffO}
\end{figure}

Based on the \Ion{O}{IV}\,/\,\Ion{O}{V} ionization equilibrium (Fig.~\ref{fig:TeffO}), we found a higher 
\Teffw{110} compared with the literature value (\Teffw{100\pm10}) for \hsb. The value of \loggw{6.0} is verified and the 
errors were reduced to $\pm$\,0.3.

\subsection{Element Abundances}
\label{sect:abund}

\begin{figure}
  \resizebox{\hsize}{!}{\includegraphics{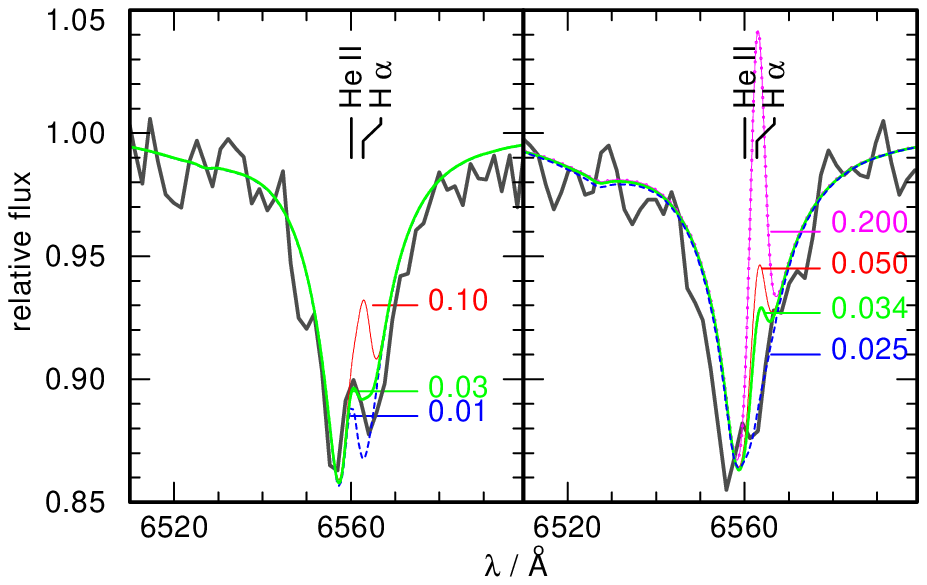}}
  \caption{Determination of the H/He ratio of \hsa (left panel) and \hsb (right). The observed H$\alpha$ and \Ion{He}{II} line
    blend is compared with models with different H/He mass ratios, as indicated by the labels.}
  \label{fig:HHe}
\end{figure}

\begin{figure}
  \resizebox{\hsize}{!}{\includegraphics{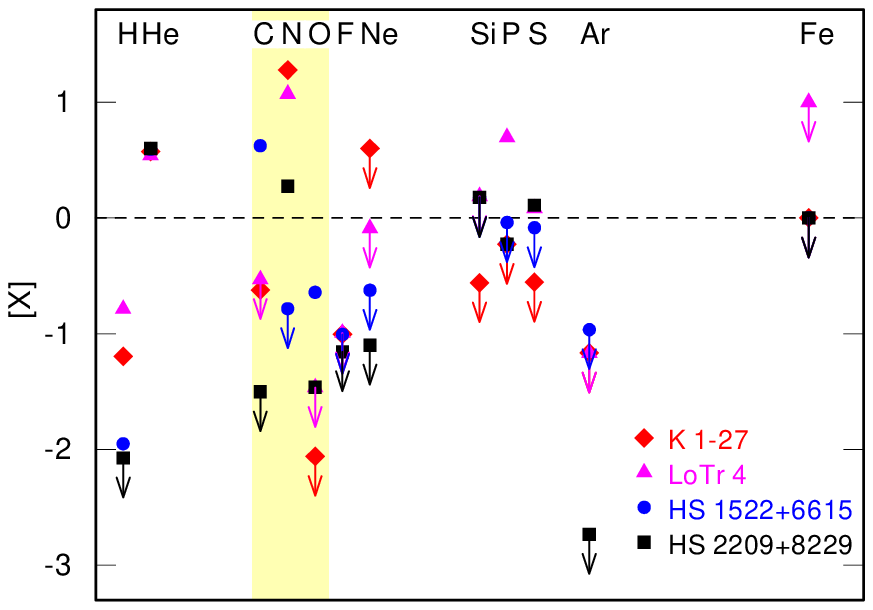}}
  \caption{Abundances of the O(He) stars (Table\,\ref{tab:parameters}). 
           The shaded (yellow) region emphasizes the CNO differences.}
  \label{fig:solar}
\end{figure}

\paragraph{{\rm{The}} H/He} 
\label{sect:hhe}

abundance ratio (by mass) was determined using 
\Ion{H}{I} $\lambda\lambda$ 6563, 4861, 4340\,\AA\ 
and \Ion{He}{II} $\lambda\lambda$ 6560, 5412, 4859, 4686, 4542, 4339\,\AA. 
It was difficult to fix because the quality of the available optical observations is poor. 
For \pnl and \pnk, we reverified the literature values of 0.5 and 0.2, respectively. 
Fig.~\ref{fig:HHe} shows that for \hsa the model with H/He\,=\,0.03 fits the observation better than 
H/He\,=\,0.10 (literature value) where H\,$\alpha$ appears in emission while for H/He\,=\,0.01 H\,$\alpha$ 
is too deep in absorption. For \hsb, we found that the upper limit of H/He ratio must be lower because of the strong 
H\,$\alpha$ emission for H/He\,$>$\,0.025 (Table\,\ref{tab:parameters}, Fig.\,\ref{fig:solar}).

\paragraph{Carbon}
\label{sect:c}

The \Ionww{C}{IV}{1548.2, 1550.8} resonance doublet in the COS spectra of all four 
stars is contaminated by the respective interstellar lines and therefore unsuited for determining the C abundance. 
In the spectra of \pnl and \hsb no other C lines were identified and we derived only 
upper limits of
$[\mathrm{C}]$\footnote{[X] denotes log (abundance by mass / solar abundance by mass) of species X. Solar abundances were
taken from \citet{asplundetal2009}.} $<-0.53$ and 
$[\mathrm{C}]<-1.50$, respectively. 
In the spectrum of \pnk, we identified \Ionww{C}{IV}{1168.8, 1168.9} and determined 
$[\mathrm{C}]=-0.62$.
For \hsa, \Ionww{C}{IV}{1168.9, 1169.0} in the COS spectrum and 
\Ionww{C}{IV}{5801.33, 5811.98} in the optical spectrum were evaluated, and
we achieve the best agreement between model and observation at $[\mathrm{C}]=0.62$.
\hsa shows the highest, supersolar C abundance (Fig.~\ref{fig:C}) while the upper limits of the 
other stars in our sample indicate subsolar abundances.

\paragraph{Nitrogen}
\label{sect:n}

The N abundance for \pnk and \pnl was determined  from the 
\Ionww{N}{V}{4603.7, 4619.7, 4933.6, 4943.2, 4945.7} emission lines. 
For \hsb, we used \Ionww{N}{V}{1616.1, 1616.4, 1619.6, 1619.7}.
As for case of \Ion{C}{IV}, the
resonance doublet \Ionww{N}{V}{1238.8, 1242.8} is blended by the respective ISM lines.
The N abundances of 
\pnk ($[\mathrm{N}]=1.28$, Fig.~\ref{fig:N}),
\pnl ($[\mathrm{N}]=1.07$), and
\hsb ($[\mathrm{N}]=2.89$) 
were found to be supersolar, in contrast to the upper limit for 
\hsa ($[\mathrm{N}]<-0.78$), which is subsolar. 

\paragraph{Oxygen}
\label{sect:o}

For \pnk and \pnl, we derived only upper limits because no O lines except the resonance doublet
\Ionww{O}{VI}{1031.9, 1037.7} were identified, which might be of ISM origin similar to the 
\Ion{C}{IV} and \Ion{N}{V} resonance doublets (Sect.\,\ref{sect:n}) 
We used the \Ionww{O}{VI}{1124.7, 1124.8} and \Ionw{O}{VI}{5290.6} to determine the upper limits 
of $[\mathrm{O}]<-2.06$ for \pnk and $[\mathrm{O}]<-1.47$ for \pnl.
For \hsa, \cite{rauchetal1998} discovered a variability of the \Ionw{O}{VI}{5290.6} emission feature. 
\citet{mickaelianetal2011} also classified \hsa as ``possible'' variable by comparing the brightness 
measurements from the Palomar Observatory Sky Survey (POSS) epochs 1 and 2. In view of its variability, 
we adjusted the O abundance to the co-added optical spectra using \Ionww{O}{VI}{3811.4, 3834.2} and 
found  $[\mathrm{O}]=-0.64$.
A precise determination of the O abundance was possible for \hsb using
\Ionw{O}{V}{1371.3}, which is prominent in the COS spectrum 
(Fig.~\ref{fig:O}). We found $[\mathrm{O}]=-1.46$ (Table\,\ref{tab:parameters}).

\paragraph{Fluorine}
\label{sect:f}

is not identified in the O(He) star spectra. We used \Ionw{F}{VI}{1139.5}
and found $[\mathrm{F}]<-1.00$ for all O(He) stars.

\paragraph{Neon}
\label{sect:ne}

We only derived upper limits using 
\Ion{Ne}{VII} $\lambda\lambda$ 1438.8, 1440.5, 1445.0\,\AA\
for \pnk, \pnl, and \hsa and 
\Ionww{Ne}{V}{1718.2, 1718.3} 
for \hsb because of its lower \Teff. 
We determined
$[\mathrm{Ne}]< 0.60$ for \pnk, 
$[\mathrm{Ne}]<-0.09$ for \pnl, 
$[\mathrm{Ne}]<-0.62$ for \hsa, and 
$[\mathrm{Ne}]<-1.10$ for \hsb.

\paragraph{Silicon}
\label{sect:si}
An upper limit for the Si abundance was determined using
\Ionw{Si}{IV}{1128.3},
\Ionw{Si}{V}{1118.8}, and
\Ionw{Si}{VI}{1130.4}.
For \pnk, it is $[\mathrm{Si}]\,<\,-0.56$. 
For \pnl, \hsa, and \hsb, we derived $[\mathrm{Si}]\,<\,0.18$.

\paragraph{Phosphorus}
\label{sect:phosphor}
we identified \Ionww{P}{V}{1118.0, 1128.0} only in the FUSE spectra  of \pnl and \hsb , 
and we determined $[\mathrm{P}] = 0.70$ and $[\mathrm{P}] = -0.76$, respectively. For \pnk and \hsa, only 
upper limits can be given, which are about solar.

\begin{table*}\centering
\caption{Photospheric parameters of the O(He) stars. Our values (R14) are compared with those given by \citet[R98]{rauchetal1998}.
         Abundances are given as logarithmic mass fraction.}
\label{tab:parameters}
\renewcommand{\tabcolsep}{1.5mm}
\begin{tabular}{rr@{.}lr@{.}lcr@{.}lr@{.}lcr@{.}lr@{.}lcr@{.}lr@{.}l} 
\hline 
\noalign{\smallskip}
             &   \multicolumn{4}{c}{\pnk}              &&  \multicolumn{4}{c}{\pnl}               && \multicolumn{4}{c}{\hsa}                && \multicolumn{4}{c}{\hsb}                             \\
\cline{2-5}
\cline{7-10}
\cline{12-15}
\cline{17-20}
\noalign{\smallskip}
             & \multicolumn{2}{c}{R14}    & 
               \multicolumn{2}{c}{R98}    && 
               \multicolumn{2}{c}{R14}    & 
               \multicolumn{2}{c}{R98}    && 
               \multicolumn{2}{c}{R14}    & 
               \multicolumn{2}{c}{R98}    && 
               \multicolumn{2}{c}{R14}    & 
               \multicolumn{2}{c}{R98}    \\ 
\noalign{\smallskip}
\hline
\noalign{\smallskip}
\Teff        & \multicolumn{4}{c}{}                    & \multicolumn{4}{c}{}                    & \multicolumn{4}{c}{}                    & \multicolumn{4}{c}{}                    \vspace{-2mm} \\
             & \multicolumn{2}{c}{$135^{+5}_{-5}$}     & 
               \multicolumn{2}{c}{$105^{+10}_{-10}$}   && 
               \multicolumn{2}{c}{$120^{+10}_{-5}$}    & 
               \multicolumn{2}{c}{$120^{+12}_{-12}$}   && 
               \multicolumn{2}{c}{$130^{+10}_{-10}$}   & 
               \multicolumn{2}{c}{$140^{+14}_{-14}$}   && 
               \multicolumn{2}{c}{$110^{+10}_{-10}$}   & 
               \multicolumn{2}{c}{$100^{+10}_{-10}$}   \vspace{-2mm} \\
kK           & \multicolumn{4}{c}{}                    & \multicolumn{4}{c}{}                    & \multicolumn{4}{c}{}                    & \multicolumn{4}{c}{}                                  \\
\hline
\noalign{\smallskip}
\logg        & \multicolumn{4}{c}{}                    & \multicolumn{4}{c}{}                    & \multicolumn{4}{c}{}                    & \multicolumn{4}{c}{}                    \vspace{-2mm} \\
             & \multicolumn{2}{c}{$6.4^{+0.2}_{-0.3}$} & 
               \multicolumn{2}{c}{$6.5^{+0.5}_{-0.5}$} && 
               \multicolumn{2}{c}{$5.8^{+0.2}_{-0.2}$} & 
               \multicolumn{2}{c}{$5.5^{+0.5}_{-0.5}$} &&
               \multicolumn{2}{c}{$5.9^{+0.2}_{-0.2}$} & 
               \multicolumn{2}{c}{$5.5^{+0.5}_{-0.5}$} && 
               \multicolumn{2}{c}{$6.0^{+0.3}_{-0.3}$} & 
               \multicolumn{2}{c}{$6.0^{+0.5}_{-0.5}$} \vspace{-2mm} \\
cm\,s$^{-2}$ & \multicolumn{4}{c}{}                    & \multicolumn{4}{c}{}                    & \multicolumn{4}{c}{}                    & \multicolumn{4}{c}{}                                  \\
\noalign{\smallskip}
\hline
\noalign{\smallskip}
H      & 
$ -1$&$33^{+0.41}_{-0.28}$    & 
$<-1$&$30$                    &&
$ -0$&$92^{+0.21}_{-0.20}$    & 
$ -0$&$90^{+0.30}_{-0.30}$    &&
$ -2$&$08^{+0.47}_{-0.52}$    & 
$ -1$&$60^{+0.30}_{-0.30}$    &&
$<-2$&$21$                    &
$<-1$&$30$                    \\
\noalign{\smallskip}
He     & 
$ -0$&$03^{+0.03}_{-0.03}$    & 
$>-0$&$02$                    &&
$ -0$&$06^{+0.02}_{-0.05}$    & 
$ -0$&$06^{+0.3}_{-0.3}$      &&
$ -0$&$009^{+0.005}_{-0.009}$ & 
$  0$&$004^{+0.3}_{-0.3}$     &&
$>-0$&$004$                   &
$>-0$&$02$                    \\
\noalign{\smallskip}
C      & 
$ -3$&$25^{+0.45}_{-0.55}$      & 
$<-1$&$82$                    &&
$<-3$&$14$                    & 
$<-1$&$92$                    &&
$ -2$&$00^{+0.68}_{-0.30}$    & 
$ -2$&$05^{+0.3}_{-0.3}$      &&
$<-4$&$13$                    &
\multicolumn{2}{c}{}          \\
\noalign{\smallskip}
N      & 
$ -1$&$88^{+0.36}_{-0.42}$    & 
$ -1$&$76^{+0.3}_{-0.3}$      &&
$ -2$&$10^{+0.60}_{-0.92}$    & 
$ -2$&$45^{+0.3}_{-0.3}$      &&
$<-3$&$94$                    & 
\multicolumn{2}{c}{}          &&
$-2$&$89^{+0.89}_{-0.50}$                    &
\multicolumn{2}{c}{}          \\
\noalign{\smallskip}
O      & 
$<-4$&$30$                    & 
\multicolumn{2}{c}{}          &&
$<-3$&$73$                    & 
$<-1$&$50$                    &&
$ -2$&$88^{+0.60}_{-0.50}$                    & 
\multicolumn{2}{c}{}          &&
$ -3$&$00^{+0.70}_{-0.70}$    &
\multicolumn{2}{c}{}          \\
\noalign{\smallskip}                                                                     
F      & 
$<-7$&$30$                    & 
\multicolumn{2}{c}{}          &&
$<-7$&$29$                    & 
\multicolumn{2}{c}{}          &&
$<-7$&$30$                    & 
\multicolumn{2}{c}{}          &&
$<-7$&$45$                    &
\multicolumn{2}{c}{}          \\
\noalign{\smallskip}                                                                    
Ne     & 
$<-2$&$30$                    & 
\multicolumn{2}{c}{}          &&
$<-2$&$99$                    & 
\multicolumn{2}{c}{}          &&
$<-3$&$53$                    & 
\multicolumn{2}{c}{}          &&
$<-4$&$00$                    &
\multicolumn{2}{c}{}          \\
\noalign{\smallskip}                                                                    
Si     & 
$<-4$&$06$                    & 
\multicolumn{2}{c}{}          &&
$<-3$&$00$                    & 
\multicolumn{2}{c}{}          &&
$<-3$&$00$                    & 
\multicolumn{2}{c}{}          &&
$<-3$&$00$                    &
\multicolumn{2}{c}{}          \\
\noalign{\smallskip}                                                                    
P      & 
$<-5$&$52$                    & 
\multicolumn{2}{c}{}          &&
$-4$&$54^{+0.40}_{-1.00}$                    & 
\multicolumn{2}{c}{}          &&
$<-5$&$20$                    & 
\multicolumn{2}{c}{}          &&
$-6$&$00^{+1.00}_{-1.00}$                    &
\multicolumn{2}{c}{}          \\
\noalign{\smallskip}                                                                    
S      & 
$<-4$&$07$                    & 
\multicolumn{2}{c}{}          &&
$ -3$&$43^{+0.61}_{-1.00}$    & 
\multicolumn{2}{c}{}          &&
$<-3$&$59$                    & 
\multicolumn{2}{c}{}          &&
$ -3$&$40^{+0.88}_{-1.27}$    &
\multicolumn{2}{c}{}          \\
\noalign{\smallskip}                                                                    
Ar     & 
$<-5$&$30$                    & 
\multicolumn{2}{c}{}          &&
$<-4$&$92$                    & 
\multicolumn{2}{c}{}          &&
$<-5$&$10$                    & 
\multicolumn{2}{c}{}          &&
$<-6$&$26$                    &
\multicolumn{2}{c}{}          \\
\noalign{\smallskip}                                                                    
Fe     & 
$<-2$&$89$                    & 
\multicolumn{2}{c}{}          &&
$<-1$&$89$                    & 
\multicolumn{2}{c}{}          &&
$<-2$&$89$                    & 
\multicolumn{2}{c}{}          &&
$<-2$&$89$                    &
\multicolumn{2}{c}{}          \\
\noalign{\smallskip}
\hline
\end{tabular}
\end{table*}

\paragraph{Sulfur}
\label{sect:s}
\Ionw{S}{VI}{1117.8} was identified in the FUSE spectra  of \pnl and \hsb
and we determined 
$[\mathrm{S}] = -0.08$ and $[\mathrm{S}] = -0.11$, respectively.
For \pnk and \hsa, only upper limits can be given, which are 0.3 times solar and solar,
respectively.

\paragraph{Argon}
\label{sect:ar}
is not identified in the O(He) star spectra. We employed \Ionw{Ar}{VII}{1535.7, 1537.1, 1537.9} and 
\Ionw{Ar}{VIII}{1164.1} and derived subsolar upper limits.

\paragraph{Iron}
\label{sect:fe}

The quality of the FUSE spectra is not sufficient to identify individual Fe lines.
Therefore, we only included the iron model atom (ionization stages {\sc v} -- {\sc ix}) 
in the line-formation calculations (i.e\@. fixed atmospheric structure) based on our final 
model and calculated NLTE occupation numbers for the atomic levels of iron. We found that the 
resulting upper limits of the Fe abundances are solar for \pnk, \hsa, 
and \hsb. For \pnl, we found an upper limit of ten times solar.

\onlfig{
\begin{figure}
  \resizebox{\hsize}{!}{\includegraphics{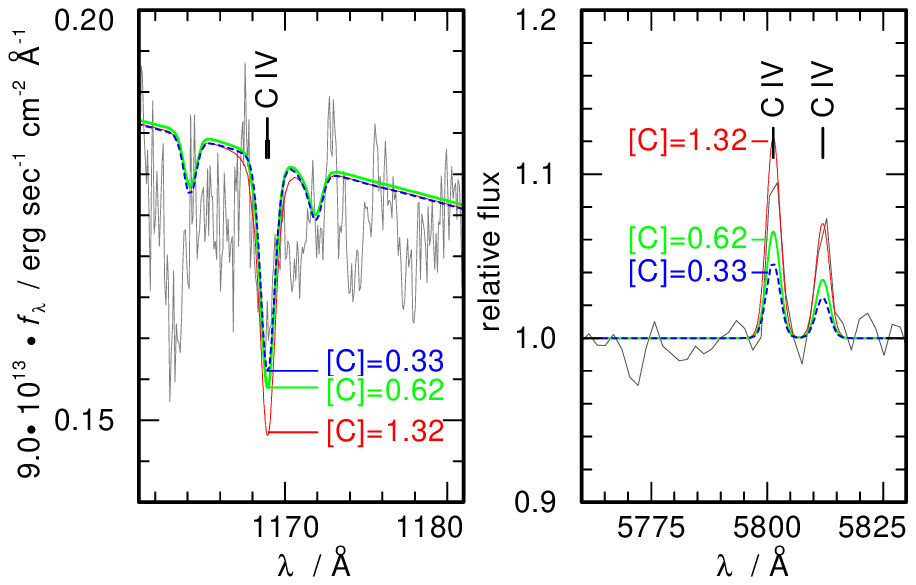}}
  \caption{Determination of the C abundance of \hsa. \Ion{C}{IV} lines are compared with models with different C abundances 
    as indicated by the labels.}
  \label{fig:C}
\end{figure}
}

\onlfig{
\begin{figure}
  \resizebox{\hsize}{!}{\includegraphics{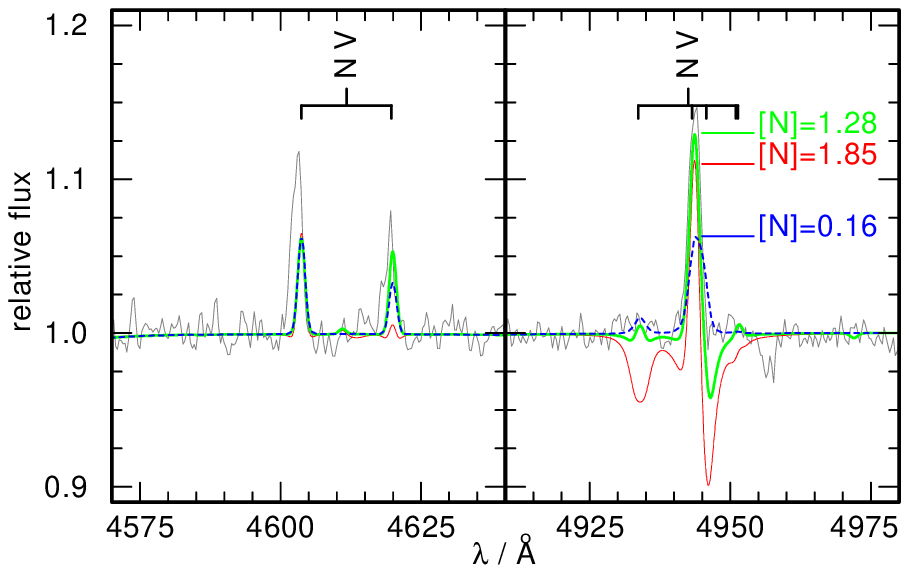}}
  \caption{Determination of the N abundance of \pnk. \Ion{N}{V} lines in the EFOSC2 observation are compared with models with different N abundances 
    as indicated by the labels.}
  \label{fig:N}
\end{figure}
}

\onlfig{
\begin{figure}
  \resizebox{\hsize}{!}{\includegraphics{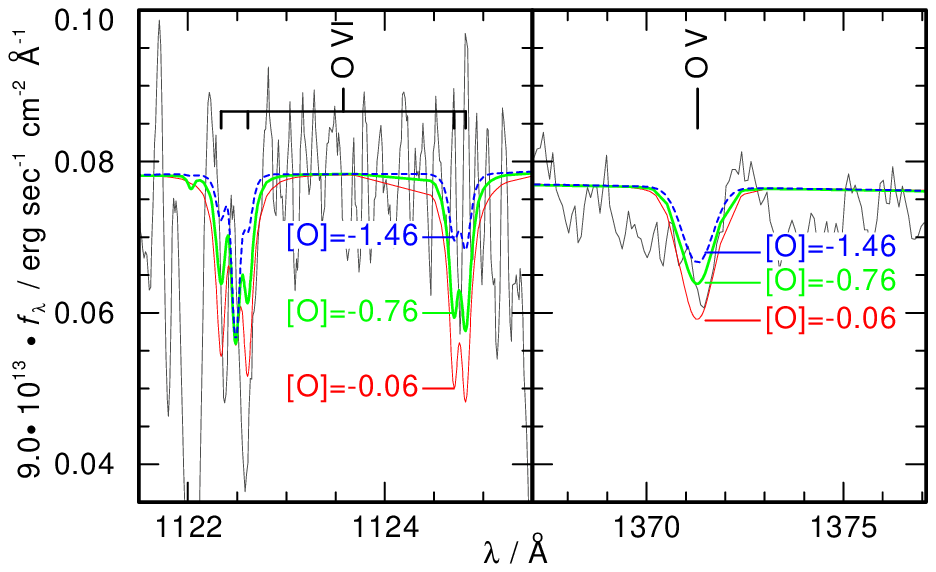}}
  \caption{Determination of the O abundance of \hsb. \Ion{O}{V} and \Ion{O}{VI} lines are compared with models with 
    different O abundances as indicated by the labels.}
  \label{fig:O}
\end{figure}
}

\subsection{Mass-loss rates}
\label{sect:massloss}

\begin{figure*}
  \resizebox{\hsize}{!}{\includegraphics{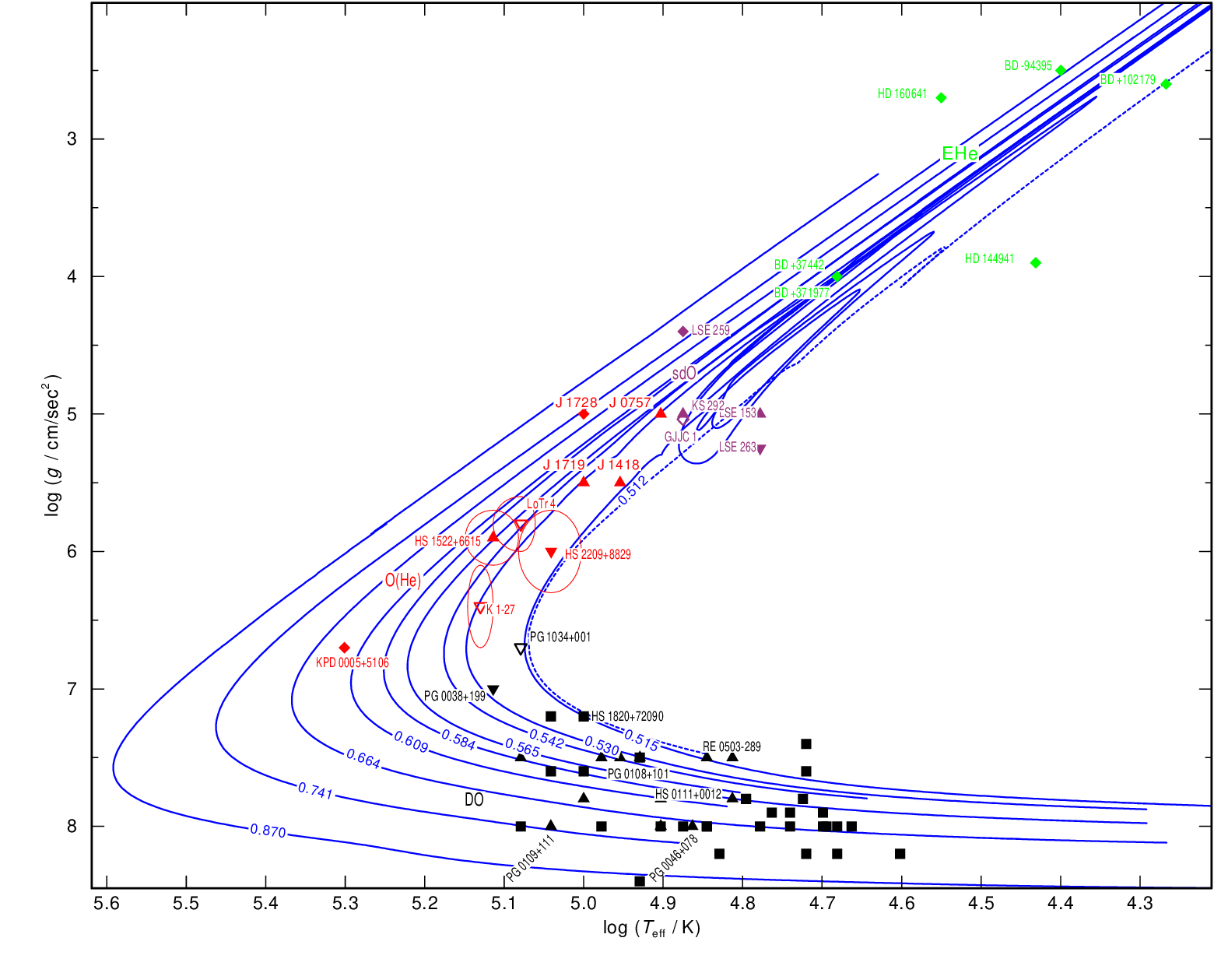}}
  \caption{Locations of EHe stars 
  \citep[green,][]{jefferyhamann2010} the luminous sdO-stars 
  (purple, \citealt{ringatPhD2013, rauchetal1998, rauchetal1991, husfeldetal1989}) and O(He) stars 
  (red, this work, \citealt{werneretal2014, wassermannetal2010}) as well as the DO WDs 
  (black, \citealt{werneretal2014, mahsereci2011, huegelmeyeretal2006, dreizlerwerner1996}) in the log \Teff\ -- \logg\ plane compared with an LTP 
  (dashed line) and VLTP post-AGB (solid lines) evolutionary tracks (labeled with stellar masses in $M_\odot$) of 
  \citet{millerbertolamialthaus2006}. Open symbols indicate CSPNe, filled ones indicate that no PN was detected. C-rich objects 
  are represented by triangles, N-rich objects by inverted triangles, C- and N-rich objects by diamonds. Squares specify objects that are neither enriched in C nor in N.}
  \label{fig:vltp}
\end{figure*}

\begin{figure*}
  \resizebox{\hsize}{!}{\includegraphics{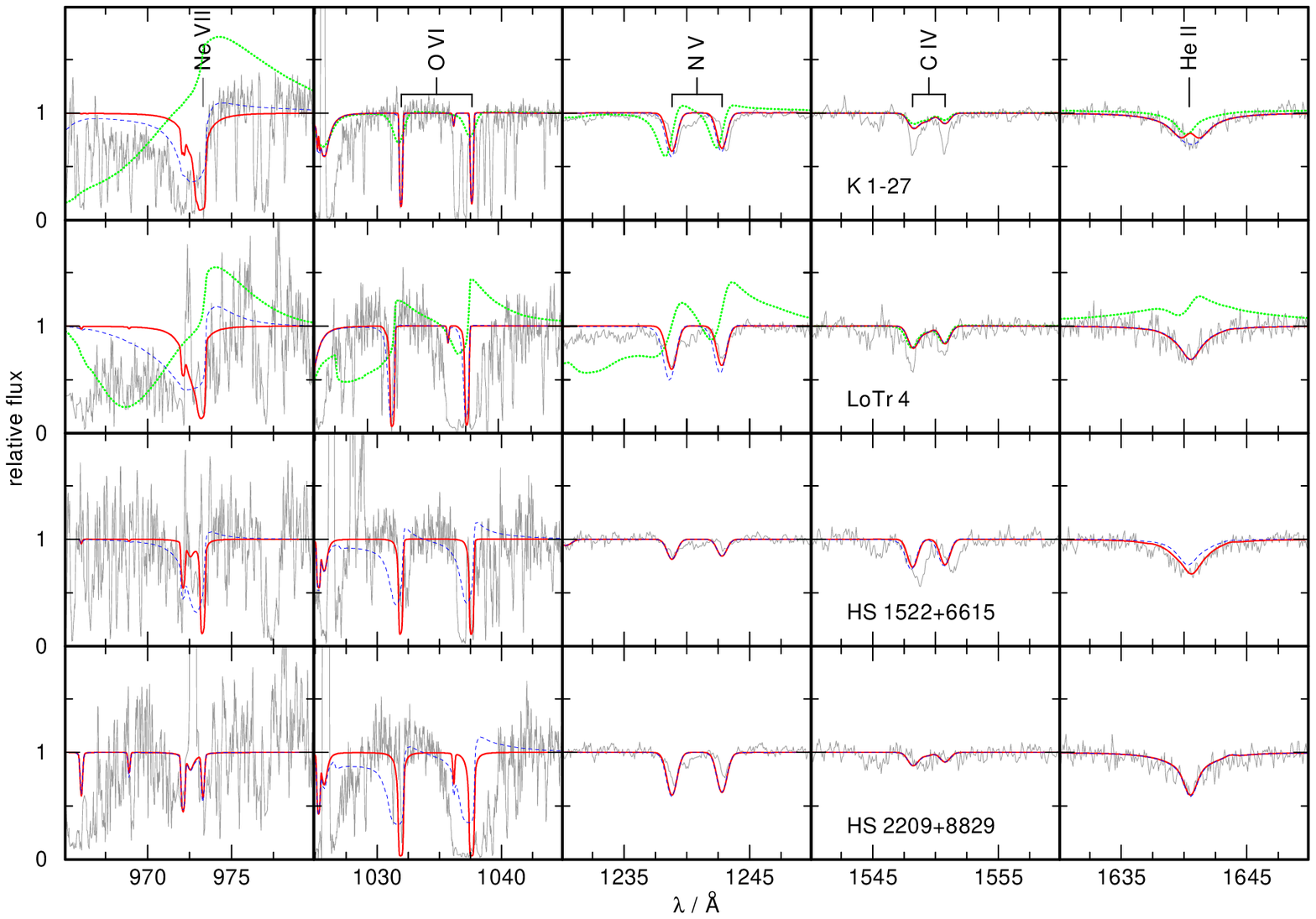}}
  \caption{Synthetic PoWR line profiles of 
           \Ion{Ne}{VII} 2p $^1$P$^\mathrm{o}$ $-$ 2p $^1$D,  
           of the resonance doublets of 
           \Ion{O}{VI}, \Ion{N}{V}, \Ion{C}{IV}, and of \Ionw{He}{II}{1640} 
           calculated with mass-loss rates from \citet[red lines]{pauldrachetal1988} and
           10\,$\times$ and 100\,$\times$ (\pnk and \pnl only) enhanced mass-loss rates (blue, dashed and green, dotted lines,
           respectively), compared with the observations of the four O(He) stars.}
  \label{fig:powr}
\end{figure*}

One aim of our COS observations was to follow up on the suggestion of \cite{millerbertolamialthaus2006} 
that O(He) stars might be post early-AGB stars. The high mass-loss rates invoked in their calculations were
not confirmed by our FUSE observations. We performed test calculations 
based on the final parameters of our analysis (Table\,\ref{tab:parameters}) 
using the Potsdam Wolf-Rayet model-atmosphere code, 
PoWR\footnote{\url{http://www.astro.physik.uni-potsdam.de/~wrh/PoWR/}},. This code solves the NLTE radiative 
transfer in a spherically expanding atmosphere simultaneously with the statistical 
equilibrium equations and accounts at the same time for energy conservation. Like in TMAP, 
iron-group line blanking is treated by means of the superlevel approach \citep{graefeneretal2002}, and a 
wind clumping in first-order approximation is taken into account \citep{hamannetal2004}. We did not 
calculate hydrodynamically consistent models, but assumed a velocity field following 
a $\beta$-law with $\beta\,=\,1$.

For our calculations, we first used the predicted mass-loss rates ($\dot M$) 
of \citet[Table\,\ref{tab:pauldrach}]{pauldrachetal1988}
using the stellar masses determined from comparison with
evolutionary tracks of VLTP stars \citep[Fig.~\ref{fig:vltp},][]{millerbertolamialthaus2006}. 
We considered the terminal wind velocities ($v_\infty$) to be 2.51 times the escape velocities of the stars 
\citep{lamers1995}. To then determine an upper limit we increased the mass-loss rate until the synthetic spectra 
no longer agreed with the observations. We exmained all strategic wind lines that are located in 
the FUSE and COS wavelength ranges (Fig.\,\ref{fig:powr}) and found that only the mass-loss rate of \pnk, which shows a weak 
P-Cygni profile at \Ionw{Ne}{VII}{973.3} in the FUSE spectrum, could be at most an order of magnitude higher than predicted 
by radiative-driven wind theory. The highest possible $\dot M$ values for all O(He) stars are given in Table\,\ref{tab:pauldrach}.

\begin{table}\centering
\caption[]{Predicted mass-loss rates of the O(He) stars interpolated from Fig.\,6a of \citet[P+88]{pauldrachetal1988}
           compared with upper limits determined in our analysis (cf\@. Fig.\,\ref{fig:powr}).}
\label{tab:pauldrach}
\begin{tabular}{cr@{.}lr@{.}l}
\hline\noalign{\smallskip}
       & \multicolumn{4}{c}{$\log (\dot M / M_\odot\,/\,\mathrm{yr})$}        \\
\noalign{\smallskip}
\cline{2-5} 
       & \multicolumn{4}{c}{}                                     \vspace{-5mm} \\
object & \multicolumn{4}{c}{}                                     \vspace{-2mm} \\
\noalign{\smallskip}
       & \multicolumn{2}{c}{P+88} & \multicolumn{2}{c}{this work}               \\
\noalign{\smallskip}
\hline
\noalign{\smallskip}
K\,1-27       & \hbox{}\hspace{5mm}$-$10&0 & \hbox{}\hspace{5mm}$\le\,-$9&0 \\ 
LoTr\,4       &                    $-$9&4 &                    $\le\,-$9&4 \\ 
HS\,1522+6615 &                    $-$8&8 &                    $\le\,-$8&8 \\
HS\,2209+8229 &                    $-$10&4 &                   $\le\,-$10&4 \\
\hline
\end{tabular}
\end{table}

\section{Stellar parameters and distances}
\label{sect:masses}

To derive the stellar parameters, we compared the position of the O(He) stars 
in the log \Teff\ -- \logg\ plane with different evolutionary tracks. 
Fig.\,\ref{fig:vltp} shows the location of the O(He) stars and related He-dominated objects compared with 
an LTP evolutionary track \citep{millerbertolamialthaus2007} and 
VLTP post-AGB tracks \citep{millerbertolamialthaus2006}. The LTP and VLTP tracks only differ at low surface 
gravities, and the masses derived from H-rich post-AGB evolutionary tracks \citep{millerbertolamialthaus2007} 
also agree well within the error limits with the VLTP masses. Therefore it is justified to derive the masses of the 
O(He) stars from these tracks although the VLTP scenario cannot be valid for these stars, as we show 
below.\\
We also derived the stellar parameters of the O(He) stars using the evolutionary tracks 
for double He-WD mergers \citep[][Fig.\,\ref{fig:evo}]{zhangetal2012a}. 
For \pnk, \pnl, and \hsb, we used the tracks from the slow-merger models because they reproduce the 
surface abundances of theses stars. For \hsa the fast-merger models were used, because they 
result in C-enriched atmospheres (Sect.\,\ref{subsect:merger}). Compared with what we found by comparison 
with VLTP calculations, luminosities and radii, which were derived from merger processes agree quite well, but 
the masses are up to $\Delta M=0.16\,M_\odot$ higher (Table~\ref{tab:mass}).

\begin{table}
\centering
\caption{Masses $M_{\mathrm{m}}$ and $M_{\mathrm{V}}$ interpolated from 
         double He-WD mergers (Fig.~\ref{fig:evo}) and
         evolutionary tracks of VLTP post-AGB stars (Fig.\,\ref{fig:vltp}),
         respectively, 
         luminosities $L$, 
         radii $R$, 
         Galactic coordinates $l$ and $b$, 
         distances $d$, and 
         height above the Galactic plane $z$
         of the O(He) stars, as well as the kinematical ages $t_{\mathrm{PN}}$ of the two PNe.}
\label{tab:mass}
\renewcommand{\tabcolsep}{1mm}
\begin{tabular}{rr@{.}lr@{.}lr@{.}lr@{.}l}
\hline
\hline
       & \multicolumn{2}{c}{}     & \multicolumn{2}{c}{}     & \multicolumn{2}{c}{HS}         & \multicolumn{2}{c}{HS}         \\
       & \multicolumn{2}{c}{\pnk} & \multicolumn{2}{c}{\pnl} & \multicolumn{2}{c}{1522+6615}  & \multicolumn{2}{c}{2209+8829}  \\
\hline
\noalign{\smallskip} 
$M_{\mathrm{m}}$\,/\,$M_\odot$      &    0&60$^{+0.08}_{-0.03}$  & 0&70$^{+0.08}_{-0.05}$  & 0&70$^{+0.05}_{-0.05}$  & 0&60$^{+0.10}_{-0.05}$ \\
\noalign{\smallskip} 
$M_{\mathrm{V}}$\,/\,$M_\odot$      & 0&53$^{+0.03}_{-0.01}$     & 0&54$^{+0.10}_{-0.01}$  & 0&57$^{+0.07}_{-0.03}$  & 0&52$^{+0.02}_{-0.01}$ \\
\noalign{\smallskip}
log $L$\,/\,$L_\odot$ &    3&2$\pm$0.4                &    3&6$\pm$0.3              &    3&7$\pm$0.2               & 3&2$\pm$0.2                    \\
\noalign{\smallskip}
$R$\,/\,$R_\odot$     &    0&08$^{+0.03}_{-0.03}$     &    0&14$^{+0.05}_{-0.03}$   &    0&14$^{+0.06}_{-0.02}$    & 0&12$^{+0.07}_{-0.04}$         \\
\noalign{\smallskip} 
$l/^{\circ}$          & 286&877                      & 291&434                   & 102&481                     & 117&849                       \\
$b/^{\circ}$          &$-$29&577                     & +19&258                   & +44&561                     & +21&545                       \\
\noalign{\smallskip} 
$d$\,/\,kpc                 & $2$&$00^{+0.63}_{-0.84}$   & $4$&$27^{+1.00}_{-1.15}$ &  $7$&$86^{+1.72}_{-2.27}$ &  $2$&$90^{+0.90}_{-1.24}$   \\ 
\noalign{\smallskip}
$z$\,/\,kpc                 & $-0$&$99^{-0.31}_{+0.41}$  & $1$&$41^{+0.33}_{-0.38}$ &  $5$&$51^{+1.21}_{-1.59}$ &  $1$&$07^{+0.33}_{-0.45}$   \\
\noalign{\smallskip}
$t_{\mathrm{PN}}$\,/\,yr  & \multicolumn{2}{c}{10867$^{+3421}_{-4553}$}  & \multicolumn{2}{c}{16184$^{+3794}_{-4360}$}  & \multicolumn{2}{c}{}  &  \multicolumn{2}{c}{}\\
\noalign{\smallskip} 
\noalign{\smallskip}
\hline 
\end{tabular}
\end{table}

Based on these masses, we calculated the distances (Table~\ref{tab:mass}) 
by using the flux calibration of \citet{heberetal1984} for $\lambda_\mathrm{eff} = 5454\,\mathrm{\AA}$,

$$d[\mathrm{pc}]=7.11 \times 10^{4} \cdot \sqrt{H_\nu\cdot M \times 10^{0.4\, m_{\mathrm{v}_0}-\log g}},$$

\noindent
with 
$m_\mathrm{V_o} = m_\mathrm{V} - 2.175 c$, $c = 1.47 E_{B-V}$, and 
the Eddington flux $H_\nu$ at $\lambda_{\rm eff}$ of our final model atmospheres. 
All O(He) stars and in particular \hsa are located far above or below the Galactic plane
(Table~\ref{tab:mass}). \pnk, \pnl, and \hsb may belong to the thick disk, 
which dominates in the region 1\,kpc $\lesssim$ $z$ $\lesssim$ 4\,kpc \citep{Kordopatis2011}. 
\hsa has a high radial velocity of $-180$\,km/s (measured by the shift of e.g\@. \Ionww{C}{IV}{1168.9, 1169.0} and 
\Ionw{He}{II}{1640.4} in the COS spectrum). Using the proper motions from the Sloan Digital Sky Survey Photometric Catalog 
(Release 9, \citealt{Ahn2012}), we calculated the space velocities to $U=95$\,km/s, $V=-116$\,km/s, and 
$W=-121$\,km/s, which are typical for halo stars \citep{Kordopatis2011}. From large height above the Galactic plane 
($z=5.51$\,kpc) of \hsa and its fast space velocities, we conclude that it belongs to the Galactic halo.

Using a typical expansion velocity of 20\,km/s for the PNe of \pnk and \pnl, we also calculated their 
kinematical ages. The results are given in Table~\ref{tab:mass}.

\section{Results and discussion}
\label{sect:discussion}

We re-analyzed all known O(He) stars based on the available spectra
and improved the determination of their properties (Tables \ref{tab:parameters}, \ref{tab:mass}).
We found that C, N, O, and Ne have a strong impact on the atmospheric structure and thus a strong 
effect on the resulting line profiles. This mostly affected \pnk, for which we found a 30\,kK higher 
\Teff\ than in literature. The COS observations allowed us to determine the N and O abundances for 
\hsb for the first time. The FUSE observation of \pnk allowed us to determine the C abundances for 
this star for the first time. We derived upper abundance limits for F, Ne, Si, P, S (for \pnl and \hsb the 
actual abundance values for P and S, measured from the FUSE spectrum), Ar, and Fe.

By examining all strategic wind-lines located in the FUSE and COS wavelength ranges, 
we found that the upper limits for mass-loss rates of the O(He) stars agree well with 
predictions by radiative-driven wind theory. Only \pnk might have a ten times higher mass-loss 
rate than predicted (only there did we find spectral signatures). From comparison with stellar-evolution 
calculations we found that the masses of the O(He) stars range from 0.52\,$M_\odot$ to 0.57\,$M_\odot$ 
considering (V)LTP tracks, or from 0.60\,$M_\odot$ to 0.70\,$M_\odot$ considering double He-WD merger 
tracks.\\

Before considering possible evolutionary scenarios, it is worth to mention, that the CNO abundance patterns 
are similar for all O(He) stars exept \hsa (Fig.\,\ref{fig:solar}), which exhibits strongly deviating CNO 
abundances. This suggests a dichotomy within the O(He) class. Interestingly, we find that other He-dominated objects 
(e.g. He-rich subdwarfs; \citealt{ringatPhD2013, Nemeth2012, naslim2010, hirsch2009, Ahmad2004, rauchetal1998, rauchetal1991, husfeldetal1989, stroeeretal2007}, 
DO WDs \citealt{werneretal2014, mahsereci2011, huegelmeyeretal2006, dreizlerwerner1996}) also 
show these differences in the CNO abundances\footnote{A list of all analyzed He-rich sdO stars, that are enriched in N 
and/or C, O(He) stars and DO WDs can be found at \url{http://astro.uni-tuebingen.de/~reindl/He}\,.}. 

Some objects which are enriched in N but not in C, some C-enriched objects do not show N, but there 
also are objects which are enriched in both. This suggests that there exist subclasses within the He-dominated 
objects. Each of them might have a different evolutionary history. From the 
He-dominated post-AGB objects (Fig.\,\ref{fig:vltp}), it seems that there is no correlation 
of the stellar mass and the C-enriched objects. There is no star among N-enriched objects
with a higher mass than 0.542\,$M_\odot$. This might be because for low-mass stars 
no third drege-up is predicted during the AGB evolution \citep{Mello2012}. 
The C- and N-enriched sdO and O(He) stars have masses greater than 0.64\,$M_\odot$.\\

\citet{rauchetal1998} already proposed the\\
\hbox{}\hspace{10mm}sdO(He) $\rightarrow$ O(He) $\rightarrow$ DO WD \\
evolutionary channel that runs parallel to the PG\,1159 evolution in the HRD. However, they were unable to present 
stringent clues on the evolution of O(He) stars. In the following we discuss different evolutionary 
channels to explain the origin of the O(He) stars, but also investigate the origin of other 
He-dominated objects to find possible progenitors and successors. First we will mention 
single-star evolutionary channels (Sect.\,\ref{sect:single}), such as enhanced mass-loss of post early-AGB stars, 
(V)LTP scenarios, and the hot-flasher scenario. In Sect.\,\ref{sect:binary} we consider to formation 
channels in binary systems, starting with the merger scenarios (Sect.\,\ref{subsect:merger}). We also discuss 
problems of this formation channel and some other possible binary formation channels such as enhanced mass-loss 
triggered by a planet or brown dwarf and the merger of a He and CO WD (Sect.\,\ref{subsect:alternative}).

\subsection{Single-star evolution}
\label{sect:single}

Assuming single-star evolution, \citet{rauchetal2009} suggested that low-mass O(He) stars might 
be post early-AGB stars that experienced an enhanced mass-loss which removed the H-rich 
envelope. In these stars, the first thermal pulse occurs after their departure from the AGB. A 
numerical experiment for a 0.512\,$M_\odot$ star by \cite{millerbertolamialthaus2006} has shown that 
an artificially increased mass-loss rate can cause the hydrogen deficiency and, in the later evolution, 
might turn the O(He) stars into PG\,1159 stars. However, to achieve that \cite{millerbertolamialthaus2006} had to 
assume $\log \dot M / (M_\odot\,/\,\mathrm{yr}) = -8$ for $\log$\,\Teff$> 3.8$. Such a strong wind is 
not predicted by radiative-driven wind theory nor is it seen in the spectra of the O(He) stars. Only 
\pnk could have a ten times higher mass-loss rate than predicted, but the derived upper limited 
($\log \dot M / (M_\odot\,/\,\mathrm{yr}) = -9$) is still too low to overcome the He-buffer and consequently 
turn the O(He) star into a PG\,1159 star. The previous mass-loss history of the O(He) stars is of course unknown and 
we can therefore only rule out PG\,1159 stars as possible successors of O(He) stars. However, the different 
H-abundances of the N-enriched O(He) stars, show a correlation with stellar mass and hence the post-AGB 
times and the remaining H in the atmospheres of these stars. While \pnl still shows 12\,\% H (by mass), the 
more evolved and lower mass \pnk only shows 5\,\% H. No H can be detected in the spectrum in the lowest mass O(He) 
star, \hsb. Although the predicted mass-loss rates are lower for stars with lower mass, their 
evolutionary timescales are much longer, and that is why the correlation of stellar mass and remaining H seems 
plausible if enhanced mass-loss is assumed.

\begin{figure*}
  \resizebox{\hsize}{!}{\includegraphics{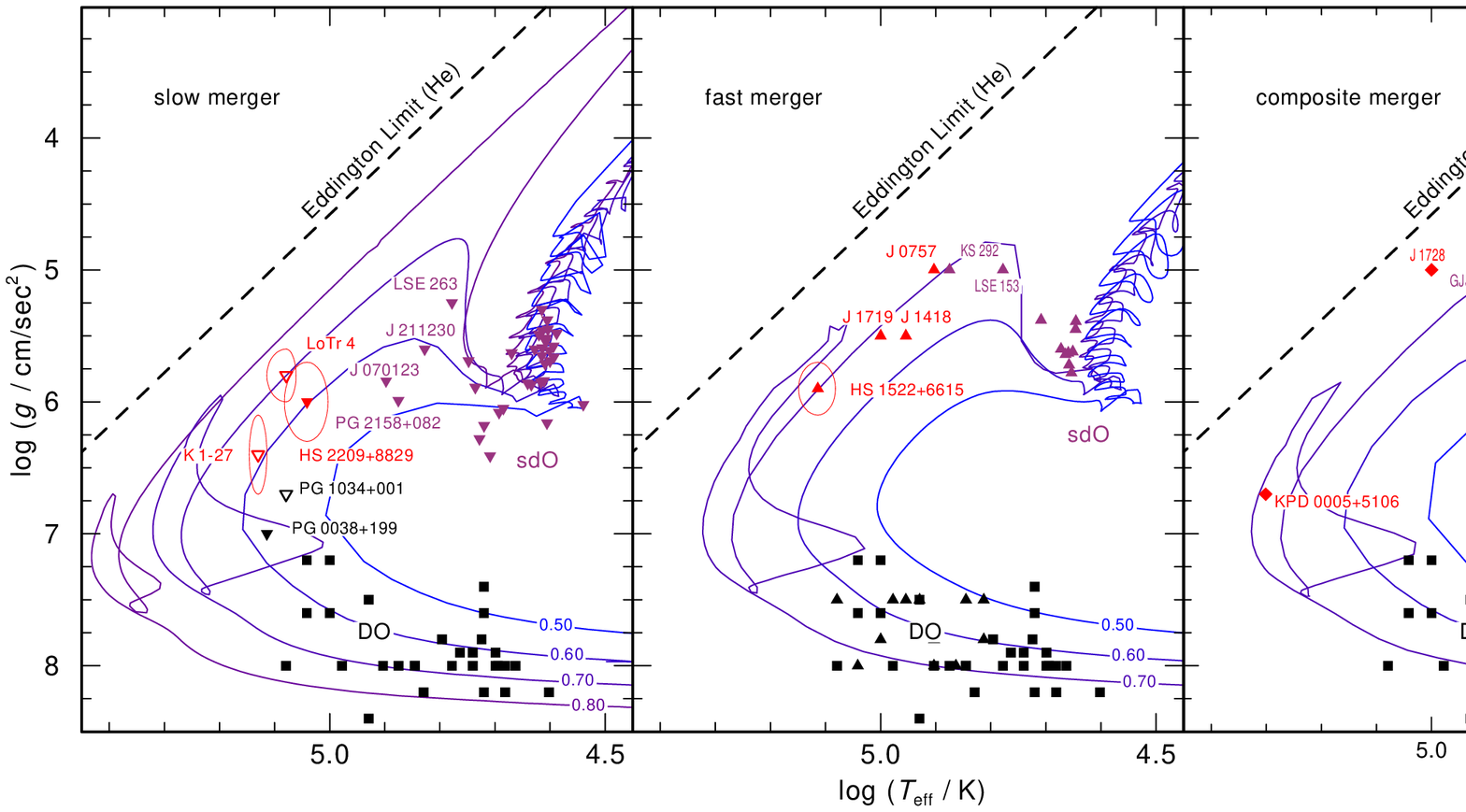}}
  \caption{Evolutionary tracks for a slow-merger (left), a fast-merger (middle) and a composite-merger (right) scenario 
    of two He-WDs, marked with the merger mass in $M_\odot$ \citep{zhangetal2012a}. 
           The locations of the sdB-, sdO-stars, (purple, \citealt{ringatPhD2013, Nemeth2012, naslim2010, hirsch2009, Ahmad2004, rauchetal1998, rauchetal1991, husfeldetal1989, stroeeretal2007}) and O(He) stars (red this work, \citealt{werneretal2014, wassermannetal2010}) as well as the DO WDs (black, \citealt{werneretal2014, mahsereci2011, huegelmeyeretal2006, dreizlerwerner1996}) are marked. 
Open symbols indicate CSPNe, filled ones that no PN was detected. C-rich objects are represented by triangles, N-rich objects by inverted triangles, C- and N-rich objects by diamonds. Squares specify objects that are neither enriched in C nor in N.}
  \label{fig:evo}
\end{figure*}

Within this scenario, good candidates for successors are C-enriched DO WDs (e.g. \wdc, \hsc, 
\citealt{dreizlerwerner1996}) for \hsa or N-enriched DO WDs (e.g. \wda, \wdb, \citealt{mahsereci2011}) 
for the three other O(He) stars. However, because of gravitational settling they might also turn into DOs, 
which do not show any C or N anymore, or even, depending on the remaining H, into DA WDs. 

Possible progenitors might be luminous helium-rich sdO-type stars, for example \lsea \citep{ringatPhD2013} for \hsa 
or \lseb \citep{ringatPhD2013} for the N-enriched O(He) stars. 

Other interesting objects that fit into this sequence, might be the [WN]-type CSPNe \ic \citep{miszalskietal2012} 
and \abe \citep{Todtetal2013, Frew2013}. Because of the strong similarity of the element abundances of [WN] stars and the 
N-enriched O(He) stars, an analog to the H-deficient, but C-rich post-AGB sequence 
[WCL] $\rightarrow$ [WCE] $\rightarrow$ PG\,1159 
\citep[e.g\@.][]{wernerherwig2006}, a second, H-deficient sequence 
[WN] $\rightarrow$ O(He) was suggested \citep[cf\@.][]{werner2012}. 
By examining \Ionw{He}{II}{4686} and \Ionw{Ne}{VII}{3890}, 
\citet{miszalskietal2012} determined \Teffw{140} for \ic. Their values for the mass and the surface 
gravity ($M = 0.6\,M_\odot$ and \loggw{6.1}) are only estimates, however. With \Teffw{140}, \ic 
would be in a similar or even later evolutionary state than the O(He) stars. \abe (\Teffw{70}), on the other hand, is in an 
earlier evolutionary stage than the O(He) stars. Still, it is necessary to clarify why \ic and \abe show a much 
stronger stellar wind (\ic: $\log \dot M / M_\odot/\mathrm{yr} = -7.7$, \abe: $\log \dot M / M_\odot/\mathrm{yr} = -6.4$) than 
the O(He) stars. We speculate that [WN] stars are O(He) stars with higher masses. According to \citet[][Fig. 6b]{pauldrachetal1988}, 
the high mass-loss rate found for \ic and \abe would correspond to $M\approx$\,0.7\,\Msol\ and $M>$1.0\,\Msol, respectively.

\cite{rauchetal2008} suggested that RCB stars might be possible progenitors of the O(He) stars. This could be true for 
the two C and N enriched O(He) stars \kwn \citep{werneretal2014} and \kpd \citep{wassermannetal2010}. In between 
evolutionary objects could be the C- and N-enriched luminous sdO stars \lsec \citep{husfeldetal1989}, \gjjc 
\citep{rauchetal1998}, and \ks \citep{rauchetal1991}.

The VLTP scenarios that were successfully applied to explain the origin of 
the H-deficient, but C-rich objects (e.g\@. PG\,1159 stars) cannot explain the origin of the 
O(He) stars because these scenarios always produce C-rich atmospheres with more than about 20\,\% of C 
by mass (in contrast, the most C-rich O(He) stars only show 3\% of C in their atmospheres). 
The relatively young kinematical age of the H-rich PNe of \pnk and \pnl (around 10\,000\,yrs) 
strongly contradict a VLTP scenario.\\

A third conceivable way for a single-star evolutionary channel is the hot-flasher scenario, 
which was invoked by \cite{millerbertolamietal2008} to explain the helium-rich sdO stars. This scenario is 
able to explain C- or N-enriched He-sdOs, but these stellar models never reach the high effective 
temperatures found for the O(He) stars. 

\subsection{Binary evolution}
\label{sect:binary}

Since none of the previously mentioned single-star formation channels seems convincing for the O(He) stars, 
binary formation channels become interesting. We discuss merger scenarios and suggest 
alternatives. 

\subsubsection{Merger scenarios}
\label{subsect:merger}

\begin{figure}
  \resizebox{\hsize}{!}{\includegraphics{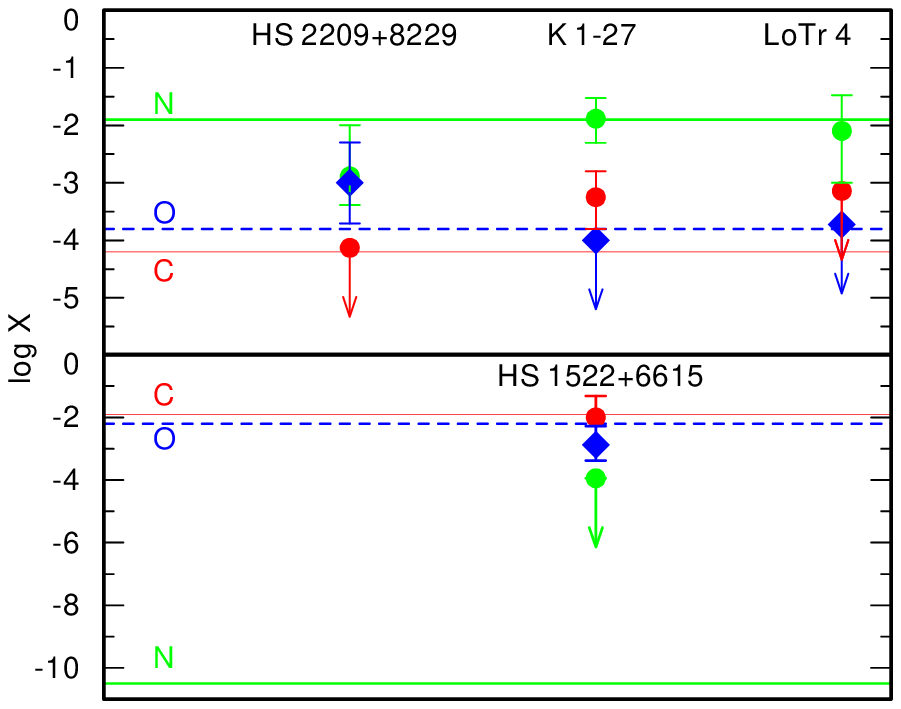}}
  \caption{Comparison of the CNO abundances of the O(He) stars 
           (N: light, green dots,
            C: dark, red dots,
            O: dark, blue rhombi,
            with those resulting form a slow 
            (top) and a fast (0.7\,\Msol\ model, bottom) merger of two He WDs 
           (N: light, green, solid line,
            C: dark, red, solid line, 
            O: dashed, blue, gray line).}
  \label{fig:fig11}
\end{figure}

As mentioned above, \citet{zhangetal2012a, zhangetal2012b} showed that in terms of \Teff, \logg, 
C, and N abundances, the origin of the three sdOs groups, but also the properties of RCB and EHe stars, 
can be explained by different double He-WD merger models. \citet{zhangetal2012b} compared their result with 
He-rich sdO stars from the sample of \cite{hirsch2009} and He-rich sdB stars from the sample 
\cite{naslim2010}. We extended this comparison to all He-rich sdB and sdO stars, that are enriched in C 
and/or N \citep{ringatPhD2013, Nemeth2012, naslim2010, 
hirsch2009, Ahmad2004, rauchetal1998, rauchetal1991, husfeldetal1989, stroeeretal2007}, the O(He) of our work, and those of 
\citet{werneretal2014, wassermannetal2010}, and the known DO WDs \citep{werneretal2014, mahsereci2011, huegelmeyeretal2006, dreizlerwerner1996}. 
We found that all of them fit this scenario. 
Figure~\ref{fig:fig11} shows the CNO abundances resulting from a slow and a fast 
merger of two He WDs according to the numerical experiments of \citet{zhangetal2012b}. These abundances are 
compared with those of the O(He) stars analyzed in our work. We found that those of the two CSPNe and \hsb can be 
explained by the slow-merger model and the CNO abundances of \hsa are reproduced by the fast-merger because of the high C abundance. 
The three O(He) stars \kwa, \kwb, and \kwc that were found by \citet{werneretal2014} also fit the fast-merger scenario. 
\kwn \citep{werneretal2014} and \kpd \citep{wassermannetal2010} would fit the composite merger model. Concerning the CNO abundances, 
the O(He) stars can be explained much better by this double He WD merger scenario than by V(LTP) scenarios. 
Fig.\,\ref{fig:evo} shows the evolutionary tracks of \citet{zhangetal2012b} for their different merger 
models. Following to their surface abundances, which are produced in a certain merger model, we marked the 
locations of the N-enriched He-dominated objects in the panel for the slow merger, C-enriched ones in the 
fast-merger panel, and C- and N-enriched objects are shown in the composite merger panel. These evolutionary 
tracks connect the subluminous sdO with the luminous sdO stars (in contrast 
to the VLTP calculations), the O(He) stars, and DO WDs.

Other points that support the merger scenario are the very low binary fraction among He-rich sdO-stars 
(4\%, \citealt{Napiwotzki2004}), O(He) stars, and DO WDs (only \cci and \ccj have been found to 
be in a binary system). A different development scenario probably applies for binary systems. 
The position of \hsa in the Galactic halo corroborates the merger scenario, because only very old stars 
are expected there. \cite{Napiwotzki2008} found that a considerable part (20\%) of the He-rich sdO stars
belongs to the halo population. 
The fact that no PN is detected around \hsa, \kwa, \kwb, and \kwc although their theoretical 
post-AGB evolutionary times would be shorter than those of \pnk or \pnl, might be explained considering that 
these objects are merger products.\\
We note that the WD+post-sdB star merger channel, proposed by \cite{justham2011} to explain He-rich sdO stars, 
can explain the O(He) stars in terms of \Teff\ and \logg as well . 

\subsubsection{Alternative scenarios}
\label{subsect:alternative}

Within the merger models mentioned above it is not possible to explain the PNe of \pnk and \pnl. Even if a 
PN would have been ejected during the merger process, it would have dissipated into the ISM a long time ago 
because the post-merger time of \pnl is about $3.7\cdot10^7$\,years. In addition, the solar compositions of the PNe 
of \pnk and \pnl and the remaining hydrogen in their atmospheres contradicts a merger 
origin of these objects. The same holds for \gjjc, \ic and \wdb. 
For these reasons \textit{common envelope (CE) scenarios} become interesting. \cite{soker2013} 
speculated that \object{R\,CrB} itself and similar RCB stars are formed from the merger of a 
WD with the core of an AGB star. The previous common-envelope ejection (CEE) then would have 
caused the H-deficiency. This scenario would also be possible for the O(He) stars. 
The merger within a CE of a CO-WD and a RG or AGB star might produce a 
similar outcome as a \textit{He- and CO-WD merger} but \textit{within a H-rich envelope}. The  
CE ejected during the merger process could then first reproduce the H-rich PNe of \pnk and \pnl and then, 
the post-merger times would be much shorter because no central helium burning phase is expected for 
a He- and CO-WD merger.\\

Another possibility for a O(He) star origin is that an \textit{RG or AGB star lost its H-rich envelope with the help 
of a planet or a brown dwarf}. If a low-mass companion enters the atmosphere of an RG or AGB star, 
it spirals inwards and transfers orbital energy and angular momentum to the envelope and, thus, 
parts or all of it are removed. The companion then either stops in a close orbit or even merges 
with the more massive star \citep{diehletal2008,soker1998}.\\

\subsection{Conclusions}
\label{sect:conclusions}

The evolutionary status of the O(He) stars, but also those of other He-dominated stars, is still not 
understood. It is most likely that they are part of a second H-deficient 
evolutionary sequence. It appears plausible that there are three subclasses within this 
He-dominated sequence. For objects that are either enriched in C or N, we propose the channels\\
\hbox{}\hspace{10mm}sdO(He) $\rightarrow$ O(He) $\rightarrow$ DO WD \\
or\\
\hbox{}\hspace{10mm}sdO(He) $\rightarrow$ O(He) $\rightarrow$ DA WD \\
if there is some remaining H. 
For C- and N-rich objects, we propose\\
\hbox{}\hspace{10mm}RCB $\rightarrow$ EHe $\rightarrow$ sdO(He) $\rightarrow$ O(He) $\rightarrow$ DO WD \\
or\\
\hbox{}\hspace{10mm}RCB $\rightarrow$ EHe $\rightarrow$ sdO(He) $\rightarrow$ O(He) $\rightarrow$ DA WD \\
if that there is some remaining H.

This He-dominated sequence has most likely various formation channels. For single-star objects, merger scenarios 
seem most promising. He-dominated CSPNe, must have formed in a different way, for example via enhanced mass-loss 
during their post-AGB evolution or a merger within a CE of a CO-WD and an RG or AGB star.

To make progress, it would be highly desirable that more O(He) stars and other He-dominated objects were discovered 
to improve the statistics. Additional quantitative investigations on the binarity of these objects and on the mass-loss rates of 
RCB, EHe, and sdO stars may help to distinguish between different formation channels. A comprehensive systematic calculation 
of evolutionary models for thermal pulse and merger scenarios is a pre-requisite for comparision with results of spectral analysis.

\begin{acknowledgements}
NR is supported by the German Research Foundation (DFG, grant WE 1312/41-1),
TR by the German Aerospace Center (DLR, grant 05\,OR\,0806).  
We thank Marcelo Miguel Miller Bertolami, Simon Jeffery, Stephan Geier, and Geoffrey Clayton for helpful discussions and comments.
This research has made use of the SIMBAD database, operated at CDS, Strasbourg, France.
This research has made use of NASA's Astrophysics Data System.
This work used the profile-fitting procedure OWENS developed by M\@. Lemoine and the FUSE French Team.
Some of the data presented in this paper were obtained from the 
Mikulski Archive for Space Telescopes (MAST). STScI is operated by the 
Association of Universities for Research in Astronomy, Inc., under NASA 
contract NAS5-26555. Support for MAST for non-HST data is provided by 
the NASA Office of Space Science via grant NNX09AF08G and by other 
grants and contracts.
\end{acknowledgements}

\bibliographystyle{aa}
\bibliography{AA_ohe} 

\end{document}